\pdfoutput=1
\RequirePackage{etex}
\documentclass[longauth]{aa}
\usepackage{graphicx}
\usepackage{txfonts}
\usepackage{tabularx}
\usepackage{booktabs}
\newcommand{\highlight}[1]

\newcommand{\wasponetwoonemuCORALIEone}{\ensuremath{38124.08^{+16.63}_{-17.27}}}
\newcommand{\wasponetwoonesigmawCORALIEone}{\ensuremath{0.45^{+7.47}_{-0.44}}}
\newcommand{\wasponetwoonemuCORALIEtwo}{\ensuremath{38336.67^{+18.54}_{-19.88}}}
\newcommand{\wasponetwoonesigmawCORALIEtwo}{\ensuremath{0.63^{+7.78}_{-0.60}}}
\newcommand{\wasponetwoonemuESPRESSOone}{\ensuremath{38167.08^{+3.44}_{-2.66}}}
\newcommand{\wasponetwoonesigmawESPRESSOone}{\ensuremath{8.73^{+3.80}_{-3.33}}}
\newcommand{\wasponetwoonemuESPRESSOtwo}{\ensuremath{38180.49^{+2.90}_{-3.25}}}
\newcommand{\wasponetwoonesigmawESPRESSOtwo}{\ensuremath{2.77^{+1.85}_{-1.06}}}
\newcommand{\wasponetwoonemuESPRESSOthree}{\ensuremath{38250.81^{+1889.35}_{-1812.77}}}
\newcommand{\wasponetwoonesigmawESPRESSOthree}{\ensuremath{38.27^{+4.30}_{-4.08}}}
\newcommand{\wasponetwoonemuESPRESSOfour}{\ensuremath{38130.07^{+38.19}_{-32.04}}}
\newcommand{\wasponetwoonesigmawESPRESSOfour}{\ensuremath{0.34^{+0.89}_{-0.29}}}
\newcommand{\wasponetwoonemuESPRESSOfive}{\ensuremath{20526.92^{+21.14}_{-21.74}}}
\newcommand{\wasponetwoonesigmawESPRESSOfive}{\ensuremath{79.73^{+11.85}_{-11.82}}}
\newcommand{\wasponetwoonemuHARPS}{\ensuremath{38342.13^{+21.53}_{-16.40}}}
\newcommand{\wasponetwoonesigmawHARPS}{\ensuremath{0.013^{+0.091}_{-0.010}}}
\newcommand{\wasponetwoonemuNIRPSone}{\ensuremath{35802.25^{+2208.72}_{-2399.00}}}
\newcommand{\wasponetwoonesigmawNIRPSone}{\ensuremath{0.37^{+8.85}_{-0.37}}}
\newcommand{\wasponetwoonemuNIRPStwo}{\ensuremath{38366.56^{+1148.79}_{-1089.14}}}
\newcommand{\wasponetwoonesigmawNIRPStwo}{\ensuremath{48.09^{+6.44}_{-5.72}}}
\newcommand{\wasponetwoonePpone}{\ensuremath{1.274924803^{+0.000000039}_{-0.000000040}}}

\newcommand{\wasponetwoonetzeropone}{\ensuremath{2458119.720725^{+0.000048}_{-0.000047}}}
\newcommand{\wasponetwooneKpone}{\ensuremath{167.97^{+5.78}_{-6.98}}}
\newcommand{\wasponetwoonerho}{\ensuremath{642.64^{+4.47}_{-4.91}}}
\newcommand{\wasponetwooneppone}{\ensuremath{0.12142^{+0.00014}_{-0.00015}}}
\newcommand{\wasponetwoonebpone}{\ensuremath{0.085^{+0.032}_{-0.038}}}
\newcommand{\wasponetwooneqoneTESS}{\ensuremath{0.223^{+0.018}_{-0.017}}}
\newcommand{\wasponetwooneqtwoTESS}{\ensuremath{0.230^{+0.024}_{-0.023}}}
\newcommand{\wasponetwooneGPsigmaCORALIEone}{\ensuremath{94.53^{+17.64}_{-15.32}}}
\newcommand{\wasponetwooneGPrhoCORALIEone}{\ensuremath{1.33^{+0.29}_{-0.30}}}
\newcommand{\wasponetwooneGPsigmaCORALIEtwo}{\ensuremath{70.84^{+24.46}_{-18.93}}}
\newcommand{\wasponetwooneGPrhoCORALIEtwo}{\ensuremath{0.84^{+0.55}_{-0.46}}}
\newcommand{\wasponetwooneGPsigmaESPRESSOthree}{\ensuremath{4870.80^{+81.67}_{-84.61}}}
\newcommand{\wasponetwooneGPrhoESPRESSOthree}{\ensuremath{0.10201^{+0.00103}_{-0.00069}}}
\newcommand{\wasponetwooneGPsigmaESPRESSOfour}{\ensuremath{102.91^{+43.93}_{-33.07}}}
\newcommand{\wasponetwooneGPrhoESPRESSOfour}{\ensuremath{1.85^{+0.69}_{-0.56}}}
\newcommand{\wasponetwooneGPsigmaHARPS}{\ensuremath{60.06^{+35.23}_{-22.29}}}
\newcommand{\wasponetwooneGPrhoHARPS}{\ensuremath{2.23^{+2.47}_{-1.09}}}
\newcommand{\wasponetwooneGPsigmaNIRPSone}{\ensuremath{4770.00^{+80.32}_{-104.19}}}
\newcommand{\wasponetwooneGPrhoNIRPSone}{\ensuremath{0.1019^{+0.0016}_{-0.0013}}}
\newcommand{\wasponetwooneGPsigmaNIRPStwo}{\ensuremath{4969.42^{+13.72}_{-16.99}}}
\newcommand{\wasponetwooneGPrhoNIRPStwo}{\ensuremath{0.10022^{+0.00025}_{-0.00024}}}
\newcommand{\wasponetwooneeccpone}{\ensuremath{0}}
\newcommand{\wasponetwooneomegapone}{\ensuremath{90}}

\newcommand{\wasponetwooneab}{\ensuremath{0.0255 \pm 0.0005}}
\newcommand{\wasponetwooneincb}{\ensuremath{88.7 \pm 0.58}}
\newcommand{\wasponetwooneTdurb}{\ensuremath{2.94 \pm 0.09}}

\newcommand{\wasponetwoonemassjb}{\ensuremath{1.1 \pm 0.06}}

\newcommand{\wasponetwooneradjb}{\ensuremath{1.72 \pm 0.04}}
\newcommand{\wasponetwoonerhoplb}{\ensuremath{0.27 \pm 0.02}}
\newcommand{\wasponetwooneteqb}{\ensuremath{2333.0 \pm 59.0}}
\newcommand{\wasponetwooneSb}{\ensuremath{897.1 \pm 38.0}}

\usepackage{placeins}
\usepackage{lipsum} % For dummy text, you can remove this line in your actual document
\usepackage{graphicx}
\usepackage{amsmath}
\usepackage{verbatim}
\usepackage{appendix}
\usepackage{hyperref}
\usepackage{ulem}
\usepackage{marvosym}
\usepackage{caption}
\usepackage{afterpage}
\usepackage[version=4]{mhchem}
\usepackage[switch]{lineno}
\usepackage[dvipsnames]{xcolor}
\usepackage{orcidlink}
\hypersetup{
    colorlinks=true,
    citecolor=blue,
    linkcolor=blue,
    urlcolor=red,
}
\begin{document} 

    \author{V. Vaulato}
    \authorrunning{Vaulato et al.} % Shorter author list for headers/footers
    %\titlerunning{Constrained atmospheric abundance ratios via high-resolution spectroscopy $\&$ orbital parameters refinement for WASP-121b}

    %\title{NIRPS transmission spectra join forces with HARPS and CRIRES+: constraining elemental abundance ratios of the ultra-hot Jupiter \hbox{WASP-121b} from a multi-instrument approach}
    %\title{Joint HARPS, NIRPS, and CRIRES+ transmission spectroscopy of the ultra-hot Jupiter \hbox{WASP-121b}: Probing elemental abundance ratios}
    %\title{Constrained atmospheric abundance ratios via joint NIRPS/HARPS and CRIRES+ transmission spectroscopy, and updated orbital parameters for the ultra-hot Jupiter WASP-121b}
   %\title{NIRPS transmission spectra meet HARPS and CRIRES+ to constrain elemental abundance ratios of the ultra-hot Jupiter \hbox{WASP-121b}}%from optical to far infrared wavelengths}
% NIRPS in synergy: from optical to far-infrared wavelenghts to constrain elemental abundance ratio
\title{Atmospheric composition and circulation of the ultra-hot Jupiter WASP-121b with joint NIRPS, HARPS and CRIRES+ transit spectroscopy}
   %\subtitle{}

   %\author{
    %Valentina Vaulato\inst{1}\thanks{Corresponding author: \textbf{\texttt{valentina.vaulato@unige.ch}}}
    %}
\author{
Valentina Vaulato\inst{1,}\thanks{Corresponding author: \textbf{\texttt{valentina.vaulato@unige.ch}}}\orcidlink{0000-0001-7329-3471},
Melissa J. Hobson\inst{1}\orcidlink{0000-0002-5945-7975},
Romain Allart\inst{2}\orcidlink{0000-0002-1199-9759},
Stefan Pelletier\inst{1,2}\orcidlink{0000-0002-8573-805X},
Joost P. Wardenier\inst{2}\orcidlink{0000-0003-3191-2486},
Hritam Chakraborty\inst{1}\orcidlink{0000-0002-5177-1898},
David Ehrenreich\inst{1,3},
Nicola Nari\inst{4,5,6},
Michal Steiner\inst{1}\orcidlink{0000-0003-3036-3585},
Xavier Dumusque\inst{1}\orcidlink{0000-0002-9332-2011},
H. Jens Hoeijmakers\inst{7}\orcidlink{0000-0001-8981-6759},
\'Etienne Artigau\inst{2,8}\orcidlink{0000-0003-3506-5667},
Fr\'ed\'erique Baron\inst{2,8}\orcidlink{0000-0002-5074-1128},
Susana C. C. Barros\inst{9,10}\orcidlink{0000-0003-2434-3625},
Bj\"orn Benneke\inst{11,2}\orcidlink{0000-0001-5578-1498},
Xavier Bonfils\inst{12}\orcidlink{0000-0001-9003-8894},
Fran\c{c}ois Bouchy\inst{1}\orcidlink{0000-0002-7613-393X},
Marta Bryan\inst{13},
Bruno L. Canto Martins\inst{14}\orcidlink{0000-0001-5578-7400},
Ryan Cloutier\inst{15}\orcidlink{0000-0001-5383-9393},
Neil J. Cook\inst{2}\orcidlink{0000-0003-4166-4121},
Nicolas B. Cowan\inst{16,17}\orcidlink{0000-0001-6129-5699},
Jose Renan De Medeiros\inst{14}\orcidlink{0000-0001-8218-1586},
Xavier Delfosse\inst{12}\orcidlink{0000-0001-5099-7978},
Elisa Delgado-Mena\inst{18,9}\orcidlink{0000-0003-4434-2195},
Ren\'e Doyon\inst{2,8}\orcidlink{0000-0001-5485-4675},
Jonay I. Gonz\'alez Hern\'andez\inst{5,6}\orcidlink{0000-0002-0264-7356},
David Lafreni\`ere\inst{2}\orcidlink{0000-0002-6780-4252},
Izan de Castro Le\~ao\inst{14}\orcidlink{0000-0001-5845-947X},
Christophe Lovis\inst{1}\orcidlink{0000-0001-7120-5837},
Lison Malo\inst{2,8}\orcidlink{0000-0002-8786-8499},
Claudio Melo\inst{19},
Lucile Mignon\inst{1,12},
Christoph Mordasini\inst{20}\orcidlink{0000-0002-1013-2811},
Francesco Pepe\inst{1}\orcidlink{0000-0002-9815-773X},
Rafael Rebolo\inst{5,6,21}\orcidlink{0000-0003-3767-7085},
Jason Rowe\inst{22},
Nuno C. Santos\inst{9,10}\orcidlink{0000-0003-4422-2919},
Damien S\'egransan\inst{1},
Alejandro Su\'arez Mascare\~no\inst{5,6}\orcidlink{0000-0002-3814-5323},
St\'ephane Udry\inst{1}\orcidlink{0000-0001-7576-6236},
Diana Valencia\inst{13}\orcidlink{0000-0003-3993-4030},
Gregg Wade\inst{23,24},
Jos\'e L. A. Aguiar\inst{14}\orcidlink{0009-0006-6577-9571},
Khaled Al Moulla\inst{9,1}\orcidlink{0000-0002-3212-5778},
Babatunde Akinsanmi\inst{1}\orcidlink{0000-0001-6519-1598},
Nicholas W. Borsato\inst{25,26}\orcidlink{0000-0002-4085-6001},
Charles Cadieux\inst{2}\orcidlink{0000-0001-9291-5555},
Yann Carteret\inst{1}\orcidlink{0000-0002-6159-6528},
Ana Rita Costa Silva\inst{9,10,1}\orcidlink{0000-0003-2245-9579},
Eduardo A. S. Cristo\inst{9}\orcidlink{0000-0001-5992-7589},
Thierry Forveille\inst{12}\orcidlink{0000-0003-0536-4607},
Yolanda G. C. Frensch\inst{1,27,}\orcidlink{0000-0003-4009-0330},
Nicole Gromek\inst{15}\orcidlink{0009-0000-1424-7694},
Monika Lendl\inst{1}\orcidlink{0000-0001-9699-1459},
Bibiana Prinoth\inst{26}\orcidlink{0000-0001-7216-4846},
Angelica Psaridi\inst{1,29,30}\orcidlink{0000-0002-4797-2419},
Atanas K. Stefanov\inst{5,6}\orcidlink{0000-0002-6059-1178},
Brian Thorsbro\inst{28,7}\orcidlink{0000-0002-5633-4400},
Drew Weisserman\inst{15}\orcidlink{0000-0002-7992-469X}
}

\institute{
\inst{1}Observatoire de Gen\`eve, D\'epartement d’Astronomie, Universit\'e de Gen\`eve, Chemin Pegasi 51, 1290 Versoix, Switzerland\\
\inst{2}Institut Trottier de recherche sur les exoplan\`etes, D\'epartement de Physique, Universit\'e de Montr\'eal, Montr\'eal, Qu\'ebec, Canada\\
\inst{3}Centre Vie dans l’Univers, Facult\'e des sciences de l’Universit\'e de Gen\`eve, Quai Ernest-Ansermet 30, 1205 Geneva, Switzerland\\
\inst{4}Light Bridges S.L., Observatorio del Teide, Carretera del Observatorio, s/n Guimar, 38500, Tenerife, Canarias, Spain\\
\inst{5}Instituto de Astrof\'isica de Canarias (IAC), Calle V\'ia L\'actea s/n, 38205 La Laguna, Tenerife, Spain\\
\inst{6}Departamento de Astrof\'isica, Universidad de La Laguna (ULL), 38206 La Laguna, Tenerife, Spain\\
\inst{7}Division of Astrophysics, Department of Physics, Lund University, Box 118, SE-22100 Lund, Sweden\\
\inst{8}Observatoire du Mont-M\'egantic, Qu\'ebec, Canada\\
\inst{9}Instituto de Astrof\'isica e Ci\^encias do Espa\c{c}o, Universidade do Porto, CAUP, Rua das Estrelas, 4150-762 Porto, Portugal\\
\inst{10}Departamento de F\'isica e Astronomia, Faculdade de Ci\^encias, Universidade do Porto, Rua do Campo Alegre, 4169-007 Porto, Portugal\\
\inst{11}Department of Earth, Planetary, and Space Sciences, University of California, Los Angeles, CA 90095, USA\\
\inst{12}Univ. Grenoble Alpes, CNRS, IPAG, F-38000 Grenoble, France\\
\inst{13}Department of Physics, University of Toronto, Toronto, ON M5S 3H4, Canada\\
\inst{14}Departamento de F\'isica Te\'orica e Experimental, Universidade Federal do Rio Grande do Norte, Campus Universit\'ario, Natal, RN, 59072-970, Brazil\\
\inst{15}Department of Physics \& Astronomy, McMaster University, 1280 Main St W, Hamilton, ON, L8S 4L8, Canada\\
\inst{16}Department of Physics, McGill University, 3600 rue University, Montr\'eal, QC, H3A 2T8, Canada\\
\inst{17}Department of Earth \& Planetary Sciences, McGill University, 3450 rue University, Montr\'eal, QC, H3A 0E8, Canada\\
\inst{18}Centro de Astrobiolog\'ia (CAB), CSIC-INTA, Camino Bajo del Castillo s/n, 28692, Villanueva de la Ca\~nada (Madrid), Spain\\
\inst{19}European Southern Observatory (ESO), Karl-Schwarzschild-Str. 2, 85748 Garching bei M\"unchen, Germany\\
\inst{20}Space Research and Planetary Sciences, Physics Institute, University of Bern, Gesellschaftsstrasse 6, 3012 Bern, Switzerland\\
\inst{21}Consejo Superior de Investigaciones Cient\'ificas (CSIC), E-28006 Madrid, Spain\\
\inst{22}Bishop's Univeristy, Dept of Physics and Astronomy, Johnson-104E, 2600 College Street, Sherbrooke, QC, Canada, J1M 1Z7, Canada\\
\inst{23}Department of Physics, Engineering Physics, and Astronomy, Queen’s University, 99 University Avenue, Kingston, ON K7L 3N6, Canada\\
\inst{24}Department of Physics and Space Science, Royal Military College of Canada, 13 General Crerar Cres., Kingston, ON K7P 2M3, Canada\\
\inst{25}School of Mathematical and Physical Sciences, Macquarie University, Sydney, NSW 2109, Australia\\
\inst{26}Lund Observatory, Division of Astrophysics, Department of Physics, Lund University, Box 118, 221 00 Lund, Sweden\\
\inst{27}European Southern Observatory (ESO), Av. Alonso de Cordova 3107,  Casilla 19001, Santiago de Chile, Chile\\
\inst{28}Universit\'e C\^ote d’Azur, Observatoire de la C\^ote d’Azur, CNRS, Laboratoire Lagrange, 06000 Nice, France\\
\inst{29}Institute of Space Sciences (ICE, CSIC), Carrer de Can Magrans S/N, Campus UAB, Cerdanyola del Valles, E-08193, Spain\\
\inst{30}Institut d’Estudis Espacials de Catalunya (IEEC), 08860 Castelldefels (Barcelona), Spain\\
\inst{*}\email{valentina.vaulato@unige.ch}
}

   \date{Received July 4$\mathrm{^{th}}$ 2025; Accepted August 26$\mathrm{^{th}}$ 2025}

    \titlerunning{Atmospheric composition and circulation of WASP-121b}
    \authorrunning{Vaulato et al.}

% \abstract{}{}{}{}{} 
% 5 {} token are mandatory
 
  \abstract
  {Ultra-hot gas giants like \hbox{WASP-121b} provide unique laboratories for exploring atmospheric chemistry and dynamics under extreme irradiation conditions. Uncovering their chemical composition and atmospheric circulation is critical for tracing planet formation pathways. Here, we present a comprehensive atmospheric characterization of \hbox{WASP-121b} using high-resolution transit spectroscopy across the optical to infrared with HARPS, NIRPS, and CRIRES+, spanning nine transit events.
  %High-resolution transmission spectroscopy, particularly when applied in synergy across multiple instruments and wavelengths, is an effective method to access the chemical and dynamical fingerprints of exoplanetary atmospheres. It enables the detection of atomic and molecular species and the inference of key chemical and dynamical properties. 
  %Here we combine optical to infrared transit observations from HARPS, NIRPS, and CRIRES+ to characterize the atmosphere of \hbox{WASP-121b}, an ultra-hot gas giant planet with extreme thermal and dynamical conditions. 
  %We analyse nine transits from HARPS, NIRPS, and CRIRES+, complemented with five TESS photometric sectors and radial velocity measurements to refine \hbox{WASP-121b}'s orbital parameters. 
  These observations are complemented with five TESS photometric sectors, two EulerCam light curves simultaneous to the HARPS and NIRPS transits, and an extensive radial velocity dataset to refine \hbox{WASP-121b}'s orbital parameters.
  A cross-correlation analysis detects iron (Fe), carbon monoxide (CO) and vanadium (V) absorption signals with $\mathrm{SNR}$ of 5.8, 5.0, 4.7, respectively. Our retrieval analysis constrains the water (H$_2$O) abundance to -6.52$_{-0.68}^{+0.49}$ dex, although its absorption signal is effectively muted by the hydride (H$^-$) continuum. We constrain the relative abundances of volatile and refractory elements -- crucial diagnostic of atmospheric chemistry, evolution, and planet formation pathways. The retrieved abundance ratios are broadly consistent with expected values of a solar composition atmosphere in chemical equilibrium, likely indicating minimal disequilibrium chemistry alterations at the probed pressures ($\mathrm{\sim10^{-4} -10^{-3}~bar}$). We update the orbital parameters of \hbox{WASP-121b} with its largest radial velocity dataset to date. By comparing orbital velocities derived from both the radial velocity analysis and the atmospheric retrieval, we determine a non-zero  velocity offset caused by atmospheric circulation, $\Delta\mathrm{K_p}=-15\pm3~\mathrm{km~s^{-1}}$ (assuming $\mathrm{M_\star=1.38\pm0.02~M_\odot}$), consistent with predictions from either drag-free or weak-drag 3D global circulation models, while cautioning the non-negligible dependence on the stellar mass assumed. These results place new constraints on the thermal structure, dynamics, and chemical inventory of \hbox{WASP-121b}, highlighting the power of multi-wavelength, high-resolution spectroscopy to probe exoplanetary atmospheres.}


   \maketitle
%
%-------------------------------------------------------------------

\section{Introduction}
\label{sec:Introduction}
%% VV Planet and stellar properties
%% Precedenti risultati in letteratura: 
%% VV Maguire+2022
%% VV Seidel+2023
%% Hoeijmakers+2024
%% Pelletier+2024
%% Smith+2024
%% VV Wardenier+2024
%% VV Seidel+2025
%% Prinoth+2025
%% High-resolution cross-correlation spectroscopy in transmission and emission
%% Retrieval framework Brogi e Gibson
%% NIRPS GTO survey
Ultra-hot Jupiters represent the hottest class of gas giant exoplanets ($\mathrm{T_{\mathrm{eq}}>2\,000~K}$), showcasing extreme atmospheric conditions driven by the proximity to their hot, early type (typically A or F) host star~\citep[e.g.][]{hellier_orbital_2009, CollierCameron+2010, west_three_2016, delrez_wasp-121_2016, gaudi_giant_2017, lund_kelt-20b_2017, anderson_wasp-189b_2018}. These planets receive an enormous stellar irradiation, orders of magnitude larger than observed in Solar-system planets, driving the thermal dissociation of molecules and the ionisation of atoms~\citep{kitzmann_peculiar_2018, parmentier_thermal_2018, hoeijmakers_atomic_2018, hoeijmakers_spectral_2019}, atmospheric circulations~\citep{arcangeli_climate_2019, seidel_hot_2019, seidel_wind_2020, seidel_into_2021, seidel+2025_w121b_dynamics_nature}, heat transport through dissociation/recombination of molecular hydrogen~\citep[H$_2$;][]{bell_increased_2018, tan_atmospheric_2019}, and condensation of metals on the nightside~\citep{ehrenreich_nightside_2020, gandhi_retrieval_2023, pelletier_vanadium_2023}. 
% The torrid temperatures activate the spectral lines (pump the strength of the lines increasing the contrast with the continuum), making ultra-hot gas giants a breeding ground to access planetary absorption and emission signatures via transmission and emission spectroscopy, respectively. 
%At such elevated temperatures, additional spectral features appear, increasing the strength of the lines relative to the continuum and making ultra-hot gas giants ideal targets for atmospheric characterization via transmission and emission spectroscopy.
At such elevated temperatures, not only do new spectral features appear, but lines already present at lower temperatures also have their amplitude and width bolstered. This overall increase in the line strength relative to the continuum makes ultra-hot gas giants ideal targets for atmospheric characterization via transmission and emission spectroscopy.

High-resolution transmission and emission spectroscopy have proven valuable methods to characterise the atmosphere of such hot planets, in tandem with the cross-correlation technique~\citep[e.g.][]{snellen_orbital_2010, brogi_signature_2012, birkby_detection_2013, hoeijmakers_atomic_2018, hoeijmakers_spectral_2019, hoeijmakers_high-resolution_2020, ehrenreich_nightside_2020, prinoth_titanium_2022, prinoth_time-resolved_2023, prinoth+2025_W121b_4UT, vaulato+2025_hydride}. The cross-correlation method has been broadly used to infer the chemical composition of planetary atmospheres, as well as their dynamics~\citep{seidel_hot_2020, seidel+2025_w121b_dynamics_nature, wardenier_phase-resolving_2024,swaetha+2025_M1b_crires+}, by resolving individual spectral lines of broad molecular bands and measuring the planet's Doppler shift at the level of $\mathrm{km~s^{-1}}$ precisions.
Unlike low-resolution spectroscopy, high-resolution data are typically self-calibrated by removing broadband flux variations and temporal changes in flux at each wavelength, thus lacking reliable information on the planet's continuum, see however~\cite{santos_broadband_2020}. Cross-correlation is then limited in the sense of being insensitive to residual broadband variations.~\cite{brogi_retrieving_2019} introduced a statistically robust atmospheric retrieval framework to overcome such limitations and access the chemical composition and temperature-pressure profile of planetary atmospheres. 

Discovered by~\cite{delrez_wasp-121_2016}, \hbox{WASP-121b} is a benchmark ultra-hot Jupiter target~\citep[$\mathrm{R_p=1.753\pm0.036~R_J}$, $\mathrm{M_p=1.157\pm0.070~M_J}$, $\mathrm{P_{orb}\sim1.27~days}$, $\mathrm{T=2358\pm52~K}$;][]{delrez_wasp-121_2016, bourrier_hot_2020, patel_empirical_2022} in an highly misaligned orbit~\citep[$\mathrm{\lambda=87.20^{+0.41}_{-0.45}~deg}$,][]{bourrier_hot_2020}~around its F6V-type star. It has been extensively observed in both emission and transmission across a broad range of wavelengths in low and high-resolution~\citep[e.g.][]{evans_detection_2016, hoeijmakers_hot_2020,sing_absolute_2024, seidel+2025_w121b_dynamics_nature, gapp+2025_W121b_JWST}.
%

%Among the latest results, 
Exospheric species such as excited hydrogen (H$_\alpha$), and ionised iron (Fe~II) and calcium (Ca~II)~\citep[firstly detected at high altitudes by][]{sing_hubble_2019, borsa_atmospheric_2021} are found to extend beyond the Roche limit that implies the planet experiences mass loss. The 4-UT VLT/ESPRESSO~\citep{pepe_espresso_2021} partial transit data obtained during the commissioning of the spectrograph~\citep{borsa_atmospheric_2021} were also used in~\cite{seidel_detection_2023} to interpret the blue-shifted sodium feature~\citep[firstly identified by ][]{hoeijmakers_hot_2020} as a signature of equatorial day-to-nightside winds crossing the evening limb. In particular,~\cite{seidel_detection_2023} characterized the secondary Na feature visible only in egress, far more offset. \cite{maguire_high-resolution_2023} constrained the relative abundances of metals by stacking high-resolution optical spectra obtained with ESPRESSO (two transits in 1-UT mode and an archival partial transit in 4-UT mode from commissioning data, the same data as used by~\cite{borsa_atmospheric_2021} and~\cite{seidel_detection_2023}), retrieving abundance ratios generally consistent with stellar values, albeit with some exceptions. Further insights into atmospheric dynamics were gained through phase-resolved infrared transmission spectra from Gemini-S/IGRINS~\citep{park_design_2014}. Carbon monoxide and H$_2$O were detected and interpreted with 3D global circulation models~\citep[GCM;][]{wardenier_phase-resolving_2024} suggesting the importance of atmospheric drag. A new 4-UT ESPRESSO dataset, completing the partial transit reported in~\cite{borsa_atmospheric_2021},~\cite{seidel_detection_2023}, and~\cite{maguire_high-resolution_2023}, allowed ~\cite{seidel+2025_w121b_dynamics_nature} to unveil an onion-layered upper atmosphere. In this configuration (see their Figure 3), atomic iron (\ion{Fe}{i}) traces sub-to-antistellar flows (day-to-night winds) at $\sim$millibar pressures; sodium (\ion{Na}{i}) probes jet streams from the morning to evening terminator at lower pressures ($\mathrm{\sim10^{-5}~bar}$), and the hydrogen ions (\ion{H}{i}) track the remarkably red- to blue-shifted signature of the jet at very high altitudes ($\mathrm{\sim10^{-6}~bar}$).

In the context of atmospheric characterisation,~\cite{hoeijmakers_mantis_2024} reported on significant detections~\citep[earlier reported by][]{hoeijmakers_hot_2020} of atomic calcium (Ca~I), vanadium (V~I), chromium (Cr~I), manganese (Mn~I), iron (Fe~I), cobalt (Co~I), and nickel (Ni~I) in ESPRESSO dayside spectra, while atomic titanium (Ti I) and titanium oxide (TiO) were notably not detected in cross-correlation. These authors surmised that Ti-bearing species condense on the nightside and remain cold-trapped there, thus inaccessible to both transmission and emission spectroscopy.~\cite{merrit+2021_W121b_inventory_UVES} used VLT/UVES~\citep{dekker+2000_uves} transit time-series to confirm robust detections of Fe~I, Cr~I, V~I, Ca~I, K~I and exospheric H~I and Ca~II, while adding a novel detection of ionised scandium (Sc~II). However,~\cite{prinoth+2025_W121b_4UT}, leveraging the high signal-to-noise ratio of 4-UT ESPRESSO transit observations, successfully detected titanium (Ti), albeit finding the signal to be notably weaker than expected. Meanwhile, titanium oxide (TiO) still remains undetected, likely due to limitations in the current line lists. Additionally, ~\cite{prinoth+2025_W121b_4UT} reported a plethora of detected species (i.e. neutral atoms like H~I, Li~I, Na~I, Mg~I, K~I, Ca~I, Ti~I, Cr~I, Mn~I, Fe~I, as well as excited atoms such as V~II, Fe~II, Co~I, Ni~I, Ba~II, Sr~I, and Sr~II) confirming and expanding upon previous literature findings~\citep{merrit+2021_W121b_inventory_UVES, silva_detection_2022, hoeijmakers_mantis_2024}. 
Optical (ESPRESSO) and infrared~\citep[VLT/CRIRES+;][]{dorn_crires_2023} data were used by~\cite{pelletier_crires_2024} to measure the volatile-to-refractory budget and C/O ratio. An analogue study was carried out in~\cite{smith_roasting_2024} using IGRINS data, retrieving overall similar properties for the dayside atmosphere of \hbox{WASP-121b}, but arriving to contrasting conclusions regarding the atmospheric volatile-to-refractory enrichment. Indeed,~\cite{pelletier_crires_2024} concluded that the atmosphere of \hbox{WASP-121b} is likely volatile-rich, while~\cite{smith_roasting_2024} preferred a super-stellar refractory-to-volatile ratio, despite both finding a $\mathrm{C/O\sim0.70}$ under the assumption of chemical equilibrium conditions. On this topic, a recent work by~\cite{evans+2025_superstellarC/O_W121b_SiO_JWST} reports on a super-stellar C/H, O/H and C/O ratios (along with robust water, carbon monoxide, and silicon monoxide detections) obtained with a single JWST observation, suggesting that pebbles and planetesimal are important in the formation pathway of giant planets.
Among the latest published results, the Near-Infrared Planet Searcher~\citep[NIRPS;][]{Bouchy+2025} has confidently detected both H$_2$O and its thermal dissociation byproduct, OH, in the thermal spectrum of WASP-121b, providing evidence for water dissociation that aligns with model predictions~\citep{bazinet_quantifying_2025}.

In this paper, we report on the first near-infrared high resolution spectroscopic NIRPS observations of WASP-121b's transit, combined with simultaneous HARPS and archival CRIRES+ transit time series. The scientific objective is to characterise the chemical composition and ongoing chemical processes in WASP-121b's atmosphere. Furthermore, we complement the spectroscopic data with high-precision radial velocity measurements to refine the orbital parameters of the system.

Section~\ref{section:Observations} describes the spectroscopic and photometric observations. Section~\ref{section:RV} presents a global fit to the RVs and photometry in order to improve the planetary parameters, and particularly the RV semi-amplitude. It includes five TESS sectors, and a total of 1261 public RV measurements from CORALIE, HARPS, ESPRESSO, and NIRPS; this is the largest RV dataset ever used to characterize WASP-121 b. 
Section~\ref{section:data_processing} illustrates the data processing for atmospheric analyses, the cross-correlation approach, and the atmospheric retrieval recipe. Section~\ref{section:results} discusses the detections of cross-correlation functions, the key retrieval results, and a comparison with previous results in the context of global circulation models, and discuss their implications.

%--------------------------------------------------------------------
\section{Observations}
\label{section:Observations}

\begin{figure}[!h]
   \centering
\includegraphics[width=\columnwidth]{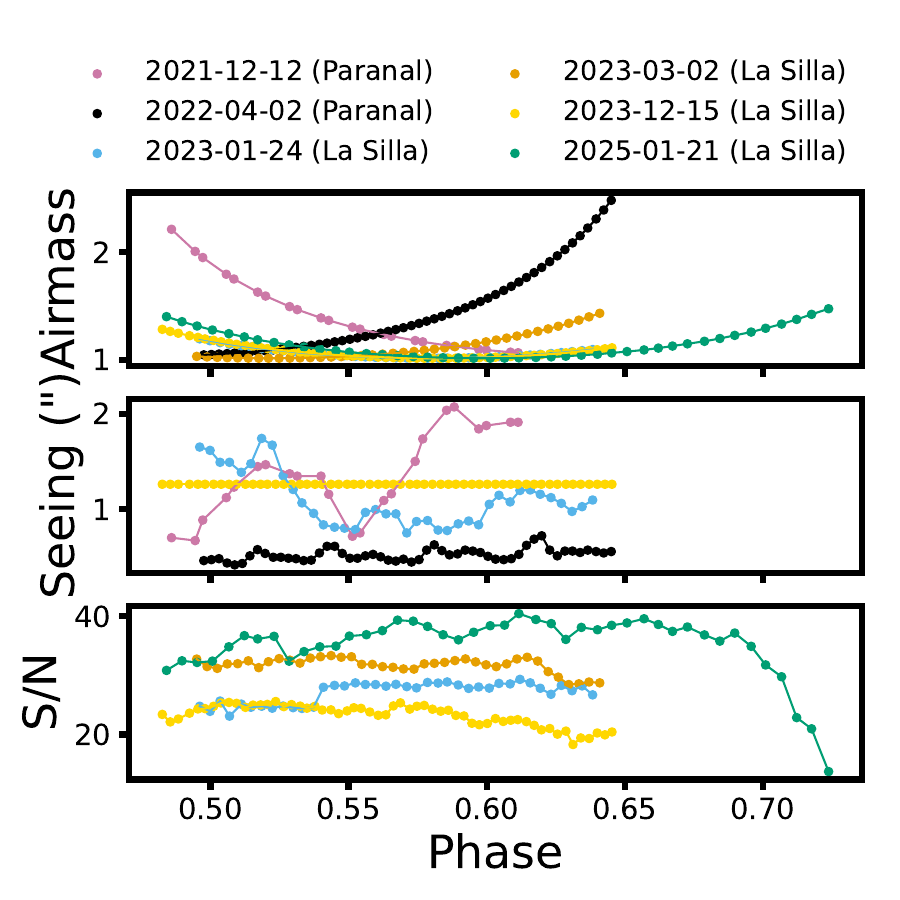}
   \caption{%Observing conditions in La Silla, Chile, as a function of phase \citep[phase = $(BJD - T_\mathrm{c})/ \mathrm{P}$, where $BJD$ is the Barycentric Julian Date, $T_\mathrm{c} = 2458926.5$ days is the mid-transit time, and $\mathrm{P}=1.27$ days is the period, both from][]{}. 
   Observing conditions as a function of phase. The blue, orange, yellow, and green points correspond to NIRPS (and simultaneous HARPS) observations, while the pink and black points represent CRIRES+ observations in the $H$- and $K$-bands, respectively. Top panel and mid panel illustrate the airmass and the average atmospheric Dimm seeing (i.e. value for the differential image monitor) evolving during the nights, respectively. The Dimm seeing is averaged for each observations (i.e. average between the start and the end of each observation). The bottom panel shows the signal-to-noise ratio (extracted from the fits header) changing during the nights.}
   \label{Figure:NightConditions}
\end{figure}
\begin{table*}[]
    \caption{Log of NIRPS, HARPS and CRIRES+ transit observations.}
    \begin{tabularx}{\textwidth}{XXX}
    \toprule
    \midrule
         & \text{Night 5} & \text{Night 6} \\
    %\midrule
    Date &  2021-12-12 & 2022-04-02 \\
    \midrule
    CRIRES+ & $K$-band  &  $H$-band \\
    \midrule
    Resolving power & 86\,000 - 92\,000 & 86\,000 - 92\,000 \\
    Number of spectra & 23 & 54 \\
    %Exp. time [s] & x & x \\
    %Number of exp. & x & x\\
    Avg. seeing ["] & 1.36 & 0.52 \\
    Avg. S/N & - & - \\
    \midrule
    \bottomrule  
    \end{tabularx}
    \label{Table:Summary_HARPSNIRPSCRIRES+_transit_obs}
    \begin{minipage}{\linewidth}
    \vspace{0.1cm}
    %\small Notes: 
    \end{minipage}
\end{table*}  
\begin{table*}[htbp]
    \begin{tabularx}{\textwidth}{XXXXX}
    \toprule
    \midrule
         & \text{Night 1} & \text{Night 2} & \text{Night 3} & \text{Night 4} \\
    %\midrule
    Date &  2023-01-24 & 2023-03-02 & 2023-12-15 & 2025-01-21 \\
    \midrule
    NIRPS & Comm. 8 & Comm. 9 & GTO & GTO \\
    \midrule
    Resolving power & 80\,000 & 80\,000 & 80\,000 & 80\,000\\
    Observing mode & HE & HA & HE & HE \\
    Number of spectra & 39 & 40 & 58 & 44 \\
    Exp. time [s] & 400 & 400 & 300 & 600 \\
    Number of exp. & 39 & 40 & 58 &  44 \\
    Avg. seeing ["] & 1.1 & 0.9 & 0.8 & 0.5 \\
    Avg. S/N order 57 $^{\star}$& 41 & 45 & 36 & 54 \\
    S/N \textit{Y} band & 32 & 38 & 28 & 40 \\
    \midrule
    HARPS & Comm. 8 & Comm. 9 & GTO & GTO \\
    \midrule
    Resolving power & 120\,000 & 120\,000 & 120\,000 & 120\,000\\
    Observing mode & - & HAM & HAM & HAM \\
    Number of spectra & - & 26 & 54 & 82 \\
    Exp. time [s] & - & 600 & 300 & 300 \\
    Number of exp. & - & 26 & 54 &  82 \\
    S/N at 550 nm & - & 30 & N.A. & 18 \\
    \midrule
    \bottomrule  
\end{tabularx}
    
    \begin{minipage}{\linewidth}
    \vspace{0.1cm}
    \small Notes: $^{\star}$In NIRPS spectra, we choose as reference the spectral order 57 centred at 16,285 $\AA$ (\textit{H} band) because it is little affected by telluric absorption lines. Night 1 and 2 belong to commissioning runs 8 and 9 (pre-GTO time), respectively.
    \end{minipage}
    \end{table*}

\subsection{Spectroscopic transit observations with NIRPS and HARPS}
\label{Spectroscopic transit observations with NIRPS and HARPS}
%HARPS DRS 3.2.5.
%Romain PI of the proposal to gather the transit dataset. CRIRES+ data slit A and slit B (B.fits are more contaminated, check runAnalysis.py file to see the set up) treated separately and eventually combined together. Having cleaned the A.fits and B.fits datasets separately, it improves the final quality of the data and thus increases the final S/N of the CO detection. It is better to proceed this way compared to using already AB combined fits files. 
We observed a total of four transits of \hbox{WASP-121b} across the disc of its relatively bright host star WASP-121~\citep[$V$=10.51, $J$=9.63;][respectively]{hog+2000_tycho2_catalog, cutri+2003_vizier_catalog} simultaneously with the high-resolution spectrographs HARPS~\citep{pepe_harps_2000} and the new fiber-fed, ultra stable and high-precision Near-Infrared Planet Searcher \citep[NIRPS;][]{Bouchy+2025}. NIRPS has been designed (i)~to expand the wavelength coverage from the optical to the $Y$, $J$, and $H$ bands, (ii)~to leverage avantgarde adaptive optics, and (iii)~to operate simultaneously with HARPS at the 3.6-m telescope in La Silla, Chile.
We observed \hbox{WASP-121b} transits on 2023-01-24 (Night 1) and 2023-03-02 (Night 2) as part of the commissioning (program ID 60.A-9109) of the NIRPS instrument, and on 2023-12-15 (Night 3) and 2025-01-21 (Night 4) as part of the ESO GTO program IDs 112.25P3.001 and 112.25P3.003, respectively (PI: F. Bouchy). Of those, three transit observations (i.e. Nights 2, 3, and 4) have been gathered simultaneously with HARPS in High Accuracy Mode (HAM) and NIRPS in High Efficiency (HE, Night 3 and 4) and High Accuracy (HA, Night 2) modes~\citep{Bouchy+2025}, whereas the first transit was observed with NIRPS only. The transit of \hbox{WASP-121b} has been partially observed by HARPS and NIRPS also on \hbox{2025-02-27}, but the dataset was discarded and not used in this work because of major technical issues (problems with the hydraulic system causing the dome vignetting the mirror) severely impacting the quality of the data (low signal-to-noise ratio). We refer the reader to Table~\ref{Table:Summary_HARPSNIRPSCRIRES+_transit_obs} and Figure~\ref{Figure:NightConditions} for a summary of the observation status. 

The transit duration (from the first to the fourth contact) is about 3 hours, and the total observing time (transit duration + baseline before ingress and after egress) was about 5 hours for all transits. All spectra were reduced by the automated HARPS and NIRPS Data Reduction Softwares \citep[DRS version 3.2.5 and 3.2.0, respectively;][]{pepe_espresso_2021}, which yield the echelle-order merged (\texttt{S1D}) and echelle-order separated (\texttt{S2D}) spectra corrected for the Barycentric Earth Radial Velocity (BERV). The starting point of the atmospheric analysis are the HARPS and NIRPS \texttt{S2D\char`_\char`BLAZE\char`_\char`A.fits} files, namely the science fiber acquired and neither blaze nor telluric corrected spectra. We thus work on a temporal series of spectra $f(\lambda, t) = f(\lambda, t)|_\mathrm{BERV}$ (i.e. provided in the barycentric rest frame).

\subsection{Spectroscopic transit observations with CRIRES+}
We supplement our optical and near-infrared observations with two CRIRES+ transit time-series (Night 5 on 2021-12-12, PI: Romain Allart, program ID: 108.22CZ;~Night 6 on 2022-04-02, PI: Brian Thorsbro, program ID 109.23D2). CRIRES+ is a high-resolution infrared slit spectrograph~\citep{follert_crires_2014, dorn_crires_2023} installed at UT3 of the Very Large Telescope in Paranal, Chile. The observations are gathered in the $K$ (Night 5) and $H$ (Night 6) spectroscopic bands, respectively. In the $K$-band, the exposures were taken in the ABBA nodding sequence (alternated position on the slit) to best subtract contributions from sky background sources, while in the $H$-band they were gathered in an unconventional AAABBBBBBAAA nodding to reduce time spent on nodding. To preserve a better time sampling and avoid additional blurring of the rapidly accelerating planetary signal, we choose not to perform our analysis on the combined AB spectra, but instead we analyse the A and B output spectra individually. Specifically, we treat the A nodding positions and the B nodding positions as separate time series for both the $H$ and $K$-band observations. That said, neither $H$ nor $K$-band data are background subtracted, but this is not worrying as it will be cared of by the principal component analysis at a later stage of the data reduction pipeline (see Section~\ref{section:data_processing} for details).

\subsection{Simultaneous photometry with EulerCam}

\begin{figure}[!h]
   \centering
\includegraphics[width=\columnwidth]{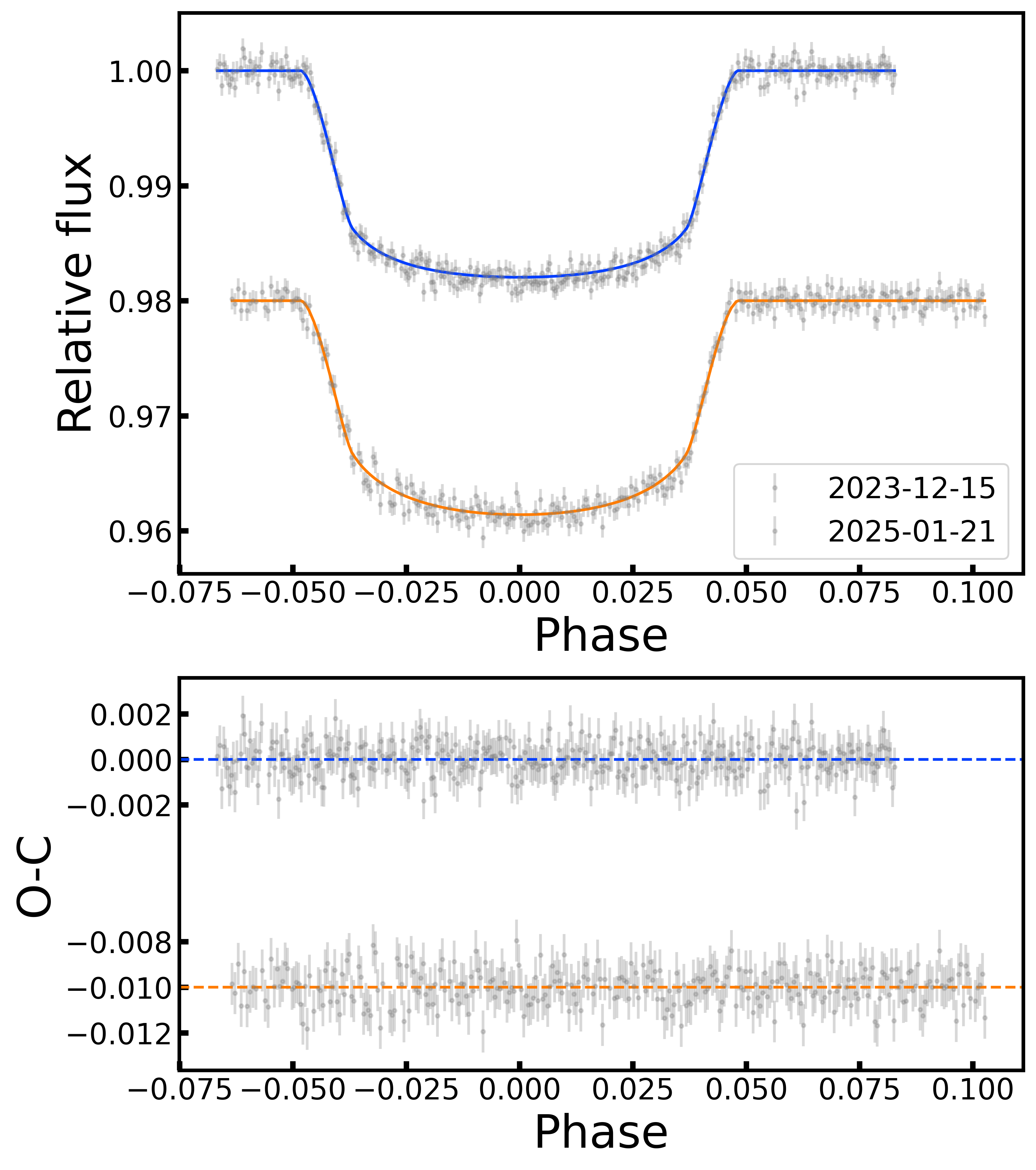}
   \caption{Simultaneous photometry with EulerCam. \textit{Top:} Detrended EulerCam light curves with the best-fit transit models overlaid on top. \textit{Bottom:} The residuals for each observation. The horizontal dashed lines indicate the continuum at zero.}
   \label{Figure:ECAM}
\end{figure}

Simultaneously with the NIRPS+HARPS observations on Nights 3 and 4 , we acquired two photometric transits using the EulerCam instrument mounted on the ESO 1.2-meter Swiss Euler telescope~\citep{Lendl+2012} on UT nights 2023-12-15 and 2025-01-21. The goal of our observations was to refine the ephemerides and to identify and mitigate the impact of any starspot occultations. The observations were taken in Geneva-$r$ and Geneva-$v$ filters with an exposure time of 30 and 45 seconds. The image reduction includes bias, flat and over-scan correction, and the aperture photometry uses the \texttt{prose} framework~\citep{garcia+2022_prose}. The optimal differential photometry is performed following~\cite{broeg+2005_differential_photometry}.  The transit light curves are modelled using the \texttt{CONAN} package \citep{Lendl+2017}. The best-fit transit model along with the detrended light curves is shown in Figure~\ref{Figure:ECAM}. The obtained ephemerides are compatible with~\cite{bourrier_hot_2020}. Thus, we choose to use the timings derived from~\cite{bourrier_hot_2020}. Furthermore, we used \texttt{PyTranSpot} \citep{juvan+2022_pyTranSpotJ, chakraborty+2024_sage} to search for any starspot occultations by simultaneously fitting a transit and spot model. In conclusion, we found no evidence of starspot contamination in the transit light curves.

%--------------------------------------------------------------------
%\section{In-depth radial velocity analysis}
\section{Refined system parameters}
\label{section:RV}
%Melissa Hobson. Opportunity to refine the ephemerides of the system (planet, star, system) using all GTO and GO ESPRESSO, HARPS, NIRPS, Euler, TESS, etc radial velocity datasets available. In particular we are interested in refining the semi-amplitude of the radial velocity of the planet (Kp).

In order to refine the orbital solution of WASP-121b, in particular the radial velocity (RV) semi-amplitudes used to calculate the value and uncertainty of $\Delta K_p$ (see Sect.~\ref{section:orbital_velocities}), we carry out a simultaneous fit of all available RV datasets and all TESS photometric data with the \texttt{juliet}\footnote{Available at \url{https://github.com/nespinoza/juliet}} software \citep{Espinoza2019juliet}. \texttt{juliet} uses the \texttt{radvel} package \citep{Fulton2018} and the \texttt{batman} package \citep{Kreidberg2015} to model RVs and transits respectively, exploring the parameter space via importance nested sampling with the \texttt{dynesty} package \citep{Speagle2020}. It employs random walk sampling with 500 live points by default. Gaussian processes (GPs) can be incorporated via the \texttt{celerite} package \citep{Foreman-Mackey2017}.

The fit incorporates ten RV datasets and five TESS sectors, which are described in Tables \ref{tab:juliet_RVs} and \ref{tab:juliet_TESS_sectors}, respectively. We remove all in-transit RVs affected by the Rossiter-McLaughlin effect, but retain the out-of transit baseline-establishing observations for these datasets, which notably allows us to incorporate three such sets of ESPRESSO data. For the TESS photometry, we use the Presearch Data Conditioning Simple Aperture Photometry (PDC-SAP) light curves \citep{Stumpe2012,Stumpe2014,Smith2012} generated by the TESS Science Processing Operations Center \citep[SPOC,][]{Jenkins2016} at NASA Ames Research Center. We obtain the light curves from the Mikulski Archive for Space Telescopes (MAST) archive. The RV data, meanwhile, are obtained from the Data \& Analysis Center for Exoplanets (DACE) platform\footnote{Available online at \url{https://dace.unige.ch}}. The specific data reduction system (DRS) versions used for each dataset, and the median RV errors, are listed in Table \ref{tab:juliet_RVs}. We treat datasets from different modes (e.g. NIRPS in high accuracy or high efficiency mode), and/or from before and after instrument upgrades (e.g. the ESPRESSO fibre link replacement, \citealt{pepe_espresso_2021}), as separate instruments for the purposes of this analysis. This allows us to account for any offsets between the datasets.

We pre-detrend the TESS sectors using a GP with the (approximate) Matern 3/2 kernel, as implemented in \texttt{celerite}, via a two-step process where we employ the out-of transit data to constrain the GP which we then apply to the in-transit data. We first mask the in-transit data in each sector, with a 5h padding of the transit window. Next, we fit a GP to the out-of-transit data in each sector, with broad log-uniform priors of $\mathcal{J}(1\times10^{-6}, 1\times10^6)$ for $\mathrm{\sigma_{GP,sector}}$ and $\mathcal{J}(1\times10^{-3}, 1\times10^{3})$ for $\mathrm{\rho_{GP,sector}}$. We also fit the flux offset $\mathrm{m_{flux,sector}}$, and jitter $\mathrm{\sigma_{w,sector}}$ for each sector using the out-of-transit data, setting the dilution factor $\mathrm{m_{dilution,sector}}$ to 1 for all sectors. We then fix the resulting parameters for the final fit, in which we use only the in-transit data.

\begin{table*}[]
\begin{center} 
\caption{RV datasets for the \texttt{juliet} modelling.}
\label{tab:juliet_RVs}
\centering 
\resizebox{\textwidth}{!}{% 
\begin{tabular}{lllllll}
\hline  \hline 
Instrument & RVs & Time span               & Mode/version & Program ID & DRS version & Median RV error [$\mathrm{m \, s^{-1}}$] \\
\hline
CORALIE1   & 31  & 11/09/2013 - 08/10/2014 & CORALIE07 & 730 & 3.4 & 19.3  \\
CORALIE2   & 19  & 18/11/2014 - 10/03/2015 & CORALIE14 & 730 & 3.8 & 41.6 \\
HARPS      & 399 & 01/01/2018 - 03/02/2024 & -  & 0100.C-0750, 112.25P3, 60.A-9109   & 3.5   & 22.8    \\
ESPRESSO2  & 10  & 01/12/2018 & HR21, pre-upgrade &    1102.C-0744  & 3.3.1 & 2.9 \\
ESPRESSO1  & 28  & 07/01/2019 & MR42, pre-upgrade  &    60.A-9128   & 3.3.1 & 5.8 \\
ESPRESSO3 & 103 & 27/01/2021 - 05/03/2021 & HR11, post-upgrade & 106.21R1, 106.21R1  & 3.3.1 & 10.4 \\
ESPRESSO4 & 298 & 12/01/2021 - 17/04/2021 & HR21, post-upgrade & 105.20BN, 106.21QM  & 3.3.1 & 5.2\\
ESPRESSO5 & 10 & 24/09/2023  & MR42, post-upgrade & 111.24J8  & 3.3.1 & 31.3 \\
NIRPS2     & 302 & 25/01/2023 - 03/02/2024 & HE  &    112.25P3, 60.A-9109 & 3.2.0 & 52.7    \\
NIRPS1     & 61  & 03/03/2023 - 06/03/2023 & HA   &   60.A-9109 & 3.2.0  & 46.2 \\
\hline
\hline
\end{tabular}
}
\end{center} 
\end{table*}

\begin{table}[]
\begin{center} 
\caption{TESS datasets for the \texttt{juliet} modelling.}
\label{tab:juliet_TESS_sectors}
\centering 
\resizebox{\columnwidth}{!}{% 
\begin{tabular}{llll}
\hline \hline
TESS sector & Camera & CCD & Time span               \\
\hline
7           & 3      & 2   & 08/01/2019 - 01/02/2019 \\
33          & 3      & 1   & 18/12/2020 - 13/01/2021 \\
34          & 3      & 2   & 14/01/2021 - 08/02/2021 \\
61          & 3      & 2   & 18/01/2023 - 12/02/2023 \\
87          & 3      & 1   & 18/12/2024 - 14/01/2025 \\
\hline
\hline
\end{tabular}
}
\end{center} 
\end{table}

We fit the following parameters in the global fit: stellar density $\rho$; the period $\mathrm{P_{b}}$, RV semi-amplitude $\mathrm{K_{b}}$ (i.e. the stellar reflex motion), time of transit $\mathrm{t_{0,b}}$, planet-to-star radius ratio $\mathrm{p_{b}}$, and impact parameter $\mathrm{b_{b}}$ for \hbox{WASP-121b}; the systemic radial velocity $\mathrm{\mu_{instrument}}$ and jitter $\mathrm{\sigma_{w,instrument}}$ for each RV instrument; and common limb-darkening parameters $\mathrm{q_{1,TESS}}$ and $\mathrm{q_{2,TESS}}$ for all TESS sectors. For the limb-darkening parameters, we used the quadratic law parametrisation of \cite{Kipping2013}, with uninformative priors of $\mathcal{U}(0,1)$. Given the very short period of the planet, we fix the eccentricity to 0. For $\mathrm{P_{b}}$, $\mathrm{t_{0,b}}$, $\mathrm{p_{b}}$, and $\mathrm{b_{b}}$ we use the values of \cite{bourrier_hot_2020} as the central point of normal priors. As we are especially interested in $\mathrm{K_{b}}$, we take a broad prior of $\mathcal{U}(0,1000)\,\mathrm{m \, s^{-1}}$. For the systemic radial velocities we set uniform priors between the minimum and maximum RV values for each instrument; for the RV jitter, we set log-uniform priors of $\mathcal{J}(0.001,100)\,\mathrm{m \, s^{-1}}$. The host star is an active fast rotator, and this activity has previously been reported to affect the RVs, necessitating their detrending \citep{delrez_wasp-121_2016, bourrier_hot_2020}. For this analysis, we also simultaneously fit a separate GP with the (approximate) Matern 3/2 kernel to each dataset except for the single-night ESPRESSO1, ESPRESSO2, and ESPRESSO5 datasets. These three datasets were excluded because their very short baselines do not allow for a good characterization of the stellar activity; we also note that preliminary test fits including a GP component for these datasets showed no improvement in their RV dispersion. To constrain the GP parameters, we first fit a GP to the FWHM activity indicator time series for each dataset, which show significant scatter and structure, similarly to \cite{Hobson2024}. We take broad uniform priors of $\mathcal{U}(1, 5000)$ for $\mathrm{\sigma_{GP}}$ and $\mathcal{U}(0.1,10)$ for $\mathrm{\rho_{GP}}$. We then use the posteriors of these fits as normal priors for the RV GPs in the full fit. 

\begin{figure*}
    \centering
    \includegraphics[width=\linewidth]{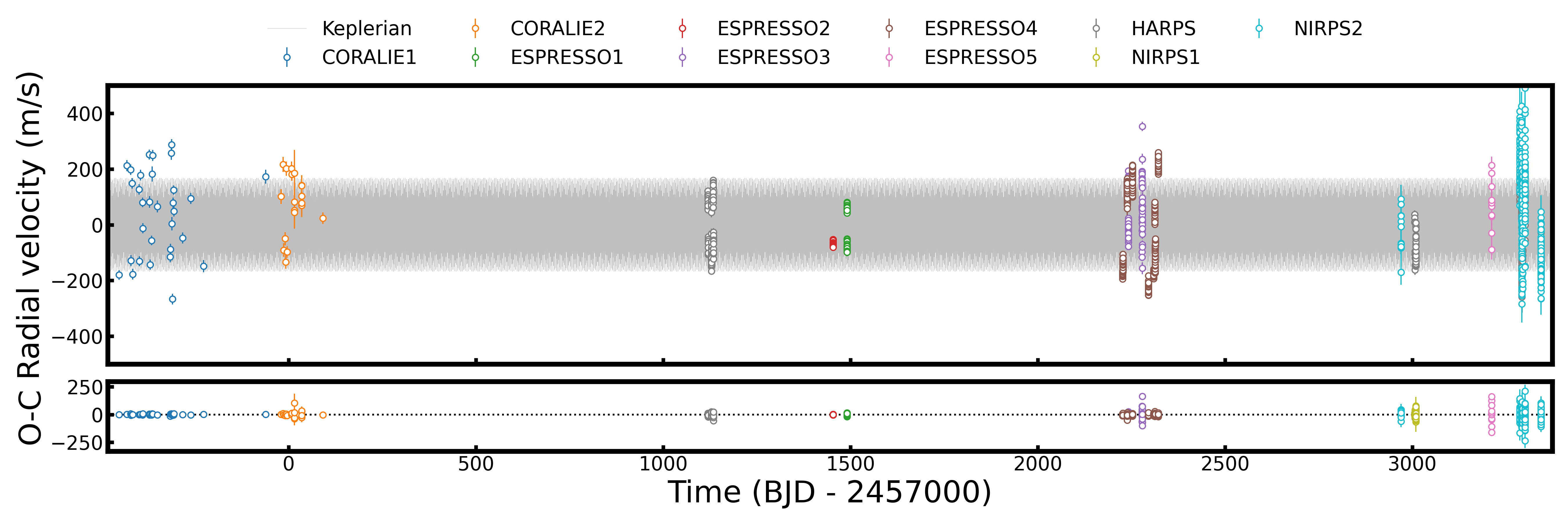}
    \caption{Top panel: RV time series (coloured dots) and best-fit Keplerian model (grey line) for the \texttt{juliet} fit of the WASP-121 public RVs. The fitted instrumental offsets $\mathrm{\mu_{instrument}}$ have been subtracted from each time series. Bottom panel: residuals of the fit.}
    \label{fig:juliet-RVs-timeseries}
\end{figure*}

\begin{figure}
    \centering
    \includegraphics[width=0.9\linewidth]{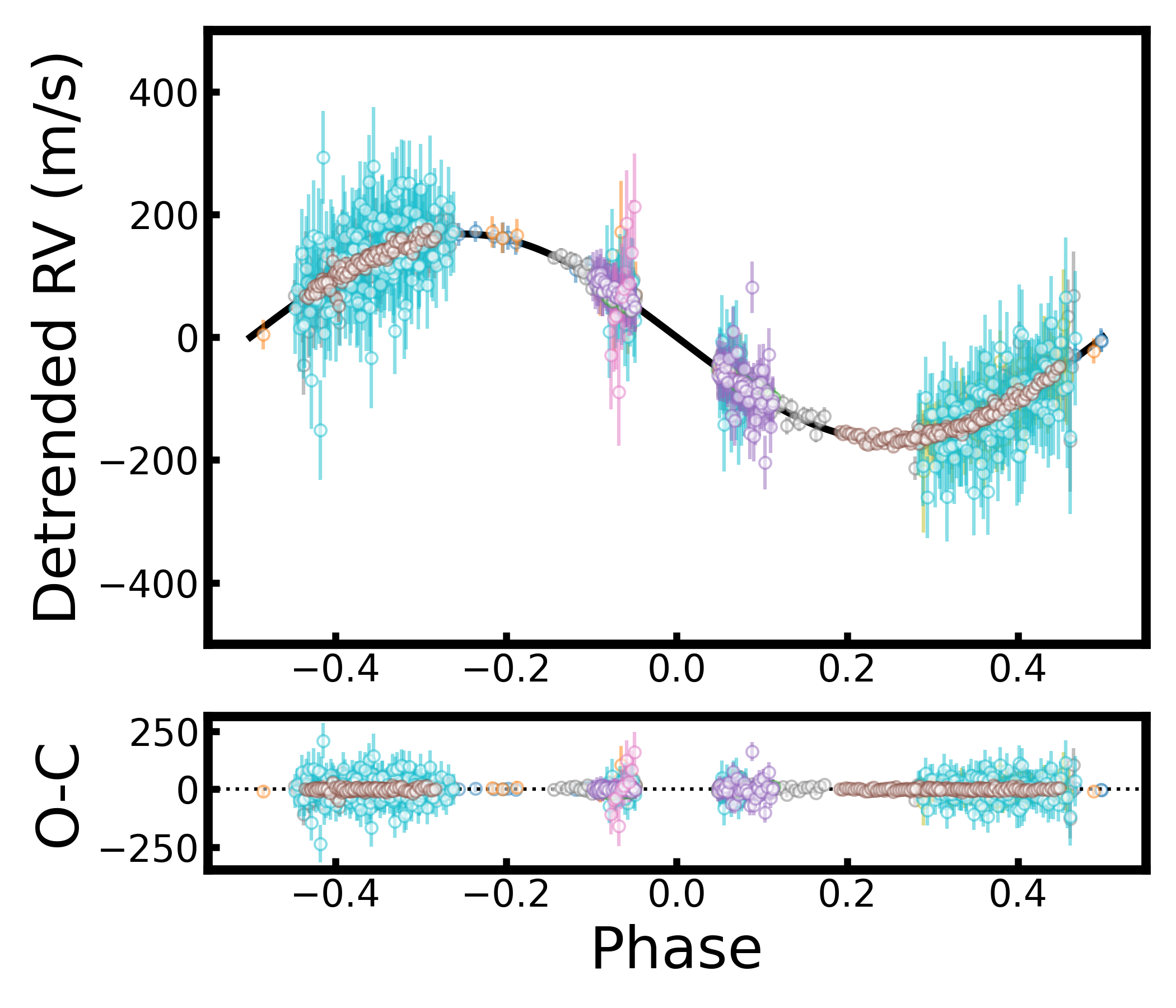}
    \caption{Top panel: phase-folded RVs (coloured dots) and best-fit Keplerian model (black line). The colours of the RV points match those in Fig. \ref{fig:juliet-RVs-timeseries}. Both the fitted instrumental offsets and the GP components (where applicable) have been subtracted from each time-series. Bottom panel: residuals of the fit.}
    \label{fig:juliet-RVs-phasefold}
    
\end{figure}

\begin{figure}
    \centering
    \includegraphics[width=0.9\linewidth]{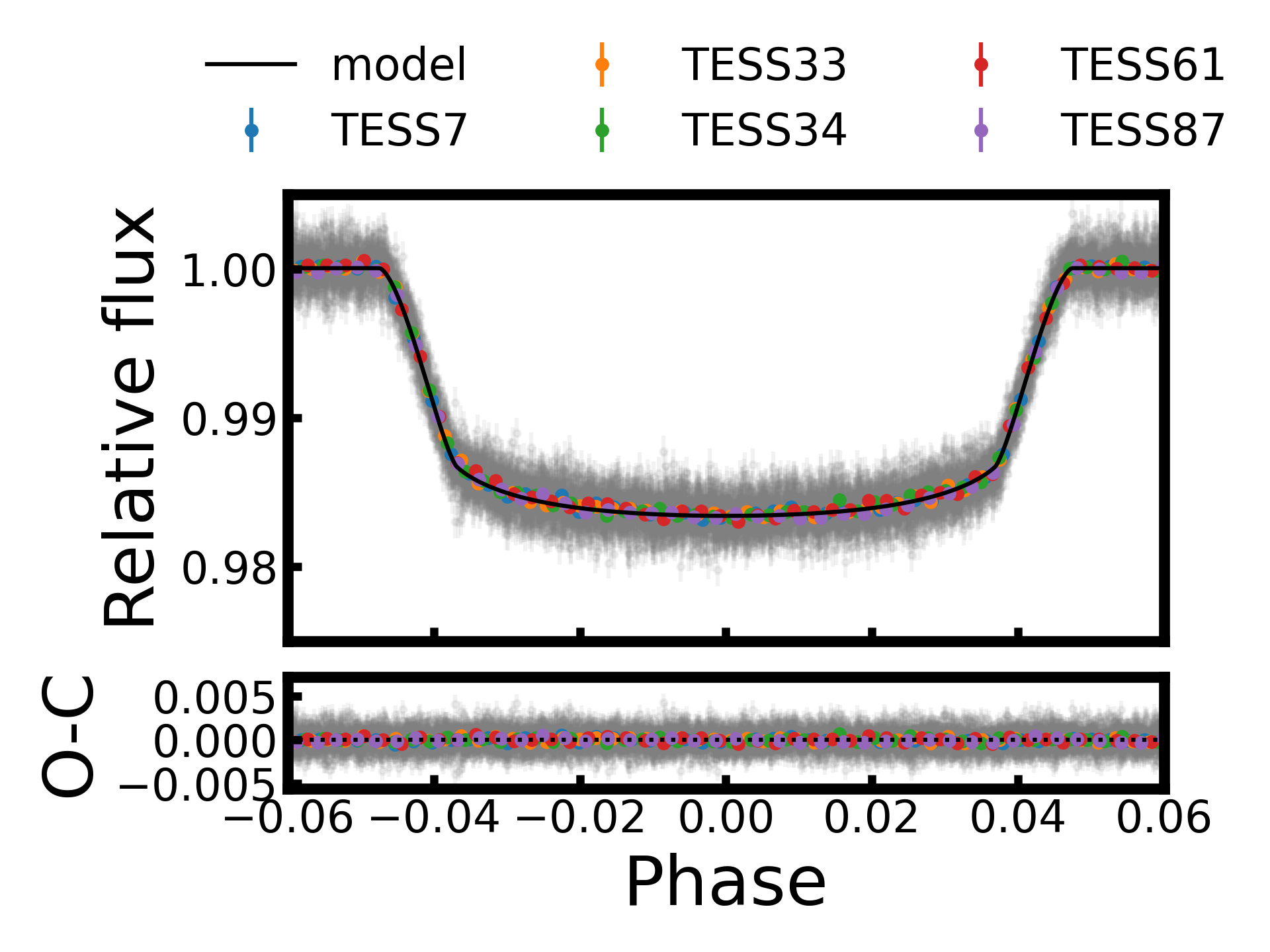}
    \caption{Top panel: stacked phase-folded TESS photometry (grey dots), binned data for each sector (coloured dots), and best-fit model (black line) for the \texttt{juliet} fit of the WASP-121 TESS photometry. Bottom panel: residuals of the fit.}
    \label{fig:juliet-TESS}
\end{figure}

The full priors and posteriors are reported in Table~\ref{tab:juliet-WASP121}. We show the RV time series and best-fit model in Figure~\ref{fig:juliet-RVs-timeseries}, the phase-folded RVs and model in Fig.~\ref{fig:juliet-RVs-phasefold} and the phase-folded TESS photometry in Fig.~\ref{fig:juliet-TESS}. We use the fitted stellar RV semi-amplitude $\mathrm{K_{b} = \wasponetwooneKpone \,m \, s^{-1}}$ to obtain a planet RV semi-amplitude of $\mathrm{K_p = 218.42 \pm 1.06 \,\mathrm{km \, s^{-1}}}$. To compute \hbox{WASP-121b}'s orbital velocity semi-amplitude, we use Equation 1 in~\cite{Torres+2010_KpEquation}, which does not depend on the planet mass. The planet mass is measured from the \texttt{juliet} RV fit, thus we prefer not to assume its value a priori to derive the orbital $\mathrm{K_p}$. To calculate the orbital planetary velocity, we assume a stellar mass $\mathrm{M_\star=1.38\pm0.02~M_\odot}$~\citep{borsa_atmospheric_2021}, an orbital inclination $\mathrm{i=88.49\pm0.16^\circ}$~\citep{bourrier_hot_2020}, and a fixed zero eccentricity. However, $\mathrm{K_p}$ is directly proportional to the stellar mass which is not easy to extract in the case of F-type stars such as WASP-121 owing to the high stellar activity and fast rotation (see Section~\ref{section:orbital_velocities} for an in depth discussion). In this work, we assume the stellar mass extracted by~\cite{borsa_atmospheric_2021} (see their Table 2, and Section 3 and 4), as they relied on the complete 1-UT transit dataset gathered with VLT/ESPRESSO.

The uncertainty on $\mathrm{K_p}$ is calculated using error propagation. In this case, $\mathrm{K_p}$ depends on the stellar mass, the orbital inclination, the stellar reflex velocity, and the orbital period. %To compute how sensitive $\mathrm{K_p}$ is to each input parameter, we use the finite difference method to approximate the partial derivatives.
To compute the uncertainty on $\mathrm{K_p}$, we use the finite difference method to approximate the partial derivatives. Thus, when including all available RV datasets (Table~\ref{tab:juliet_RVs}, Figure~\ref{fig:juliet-RVs-phasefold}), we improve the uncertainty on both the stellar reflex motion $\mathrm{K_{b}}$ and the computed semi-amplitude orbital velocity of \hbox{WASP-121b} with respect to~\cite{bourrier_hot_2020} (see their Table 1).

\begin{table}[pht] 
\begin{center} 
\caption{Prior and posterior planetary parameter distributions obtained with \texttt{juliet} for WASP-121.} 
\label{tab:juliet-WASP121} 
\centering 
\resizebox{\columnwidth}{!}{% 
\begin{tabular}{lll} 
\hline  \hline 
Parameter & Prior & Posterior \\ 
\hline 
$\mathrm{\mu_{CORALIE1}}$ \dotfill [$\mathrm{m \, s^{-1}}$]& $\mathcal{U}(18928.81,76823.52)$ & \wasponetwoonemuCORALIEone \\
$\mathrm{\sigma_{w,CORALIE1}}$ \dotfill [$\mathrm{m \, s^{-1}}$]& $\mathcal{J}(0.001,100)$ & \wasponetwoonesigmawCORALIEone \\
$\mathrm{\mu_{CORALIE2}}$ \dotfill [$\mathrm{m \, s^{-1}}$]& $\mathcal{U}(19047.03,77104.58)$ & \wasponetwoonemuCORALIEtwo \\
$\mathrm{\sigma_{w,CORALIE2}}$ \dotfill [$\mathrm{m \, s^{-1}}$]& $\mathcal{J}(0.001,100)$ & \wasponetwoonesigmawCORALIEtwo \\
$\mathrm{\mu_{ESPRESSO1}}$ \dotfill [$\mathrm{m \, s^{-1}}$]& $\mathcal{U}(19032.28,76495.73)$ & \wasponetwoonemuESPRESSOone \\
$\mathrm{\sigma_{w,ESPRESSO1}}$ \dotfill [$\mathrm{m \, s^{-1}}$]& $\mathcal{J}(0.001,100)$ & \wasponetwoonesigmawESPRESSOone \\
$\mathrm{\mu_{ESPRESSO2}}$ \dotfill [$\mathrm{m \, s^{-1}}$]& $\mathcal{U}(19041.41,76296.34)$ & \wasponetwoonemuESPRESSOtwo \\
$\mathrm{\sigma_{w,ESPRESSO2}}$ \dotfill [$\mathrm{m \, s^{-1}}$]& $\mathcal{J}(0.001,100)$ & \wasponetwoonesigmawESPRESSOtwo \\
$\mathrm{\mu_{ESPRESSO3}}$ \dotfill [$\mathrm{m \, s^{-1}}$]& $\mathcal{U}(18944.17,76794.85)$ & \wasponetwoonemuESPRESSOthree \\
$\mathrm{\sigma_{w,ESPRESSO3}}$ \dotfill [$\mathrm{m \, s^{-1}}$]& $\mathcal{J}(0.001,100)$ & \wasponetwoonesigmawESPRESSOthree \\
$\mathrm{\mu_{ESPRESSO4}}$ \dotfill [$\mathrm{m \, s^{-1}}$]& $\mathcal{U}(18938.14,76776.31)$ & \wasponetwoonemuESPRESSOfour \\
$\mathrm{\sigma_{w,ESPRESSO4}}$ \dotfill [$\mathrm{m \, s^{-1}}$]& $\mathcal{J}(0.001,100)$ & \wasponetwoonesigmawESPRESSOfour \\
$\mathrm{\mu_{ESPRESSO5}}$ \dotfill [$\mathrm{m \, s^{-1}}$]& $\mathcal{U}(10219.46,41957.50)$ & \wasponetwoonemuESPRESSOfive \\
$\mathrm{\sigma_{w,ESPRESSO5}}$ \dotfill [$\mathrm{m \, s^{-1}}$]& $\mathcal{J}(0.001,100)$ & \wasponetwoonesigmawESPRESSOfive \\
$\mathrm{\mu_{HARPS}}$ \dotfill [$\mathrm{m \, s^{-1}}$]& $\mathcal{U}(19041.74,77090.00)$ & \wasponetwoonemuHARPS \\
$\mathrm{\sigma_{w,HARPS}}$ \dotfill [$\mathrm{m \, s^{-1}}$]& $\mathcal{J}(0.001,100)$ & \wasponetwoonesigmawHARPS \\
$\mathrm{\mu_{NIRPS1}}$ \dotfill [$\mathrm{m \, s^{-1}}$]& $\mathcal{U}(19122.74,77361.06)$ & \wasponetwoonemuNIRPSone \\
$\mathrm{\sigma_{w,NIRPS1}}$ \dotfill [$\mathrm{m \, s^{-1}}$]& $\mathcal{J}(0.001,100)$ & \wasponetwoonesigmawNIRPSone \\
$\mathrm{\mu_{NIRPS2}}$ \dotfill [$\mathrm{m \, s^{-1}}$]& $\mathcal{U}(19053.58,77764.47)$ & \wasponetwoonemuNIRPStwo \\
$\mathrm{\sigma_{w,NIRPS2}}$ \dotfill [$\mathrm{m \, s^{-1}}$]& $\mathcal{J}(0.001,100)$ & \wasponetwoonesigmawNIRPStwo \\
$\mathrm{P_{b}}$ \dotfill [d]& $\mathcal{N}(1.27492504,0.001)$ & \wasponetwoonePpone \\
$\mathrm{t_{0,p1}}$ \dotfill [BJD]& $\mathcal{N}(2458119.72074,0.01)$ & \wasponetwoonetzeropone \\
$\mathrm{K_{b}}$ \dotfill [$\mathrm{m \, s^{-1}}$]& $\mathcal{U}(0,1000)$ & \wasponetwooneKpone \\
$\mathrm{\rho_{}}$ \dotfill [$\mathrm{kg \, m^{-3}}$]& $\mathcal{J}(100,10000)$ & \wasponetwoonerho \\
$\mathrm{p_{b}}$ \dotfill & $\mathcal{N}(0.12355,0.1)$ & \wasponetwooneppone \\
$\mathrm{b_{b}}$ \dotfill & $\mathcal{N}(0.1,0.1)$ & \wasponetwoonebpone \\
$\mathrm{q_{1,TESS}}$ \dotfill & $\mathcal{U}(0,1)$ & \wasponetwooneqoneTESS \\
$\mathrm{q_{2,TESS}}$ \dotfill & $\mathcal{U}(0,1)$ & \wasponetwooneqtwoTESS \\
$\mathrm{\sigma_{GP,CORALIE1}}$ \dotfill & $\mathcal{N}(428.64,82.00)$ & \wasponetwooneGPsigmaCORALIEone \\
$\mathrm{\rho_{GP,CORALIE1}}$ \dotfill & $\mathcal{N}(1.60,0.45)$ & \wasponetwooneGPrhoCORALIEone \\
$\mathrm{\sigma_{GP,CORALIE2}}$ \dotfill & $\mathcal{N}(238.04,93.20)$ & \wasponetwooneGPsigmaCORALIEtwo \\
$\mathrm{\rho_{GP,CORALIE2}}$ \dotfill & $\mathcal{N}(0.86,1.29)$ & \wasponetwooneGPrhoCORALIEtwo \\
$\mathrm{\sigma_{GP,ESPRESSO3}}$ \dotfill & $\mathcal{N}(4935.53,96.44)$ & \wasponetwooneGPsigmaESPRESSOthree \\
$\mathrm{\rho_{GP,ESPRESSO3}}$ \dotfill & $\mathcal{N}(0.1008,0.0014)$ & \wasponetwooneGPrhoESPRESSOthree \\
$\mathrm{\sigma_{GP,ESPRESSO4}}$ \dotfill & $\mathcal{N}(831.27,328.02)$ & \wasponetwooneGPsigmaESPRESSOfour \\
$\mathrm{\rho_{GP,ESPRESSO4}}$ \dotfill & $\mathcal{N}(1.98,0.92)$ & \wasponetwooneGPrhoESPRESSOfour \\
$\mathrm{\sigma_{GP,HARPS}}$ \dotfill & $\mathcal{N}(1035.30,367.75)$ & \wasponetwooneGPsigmaHARPS \\
$\mathrm{\rho_{GP,HARPS}}$ \dotfill & $\mathcal{N}(6.38,2.07)$ & \wasponetwooneGPrhoHARPS \\
$\mathrm{\sigma_{GP,NIRPS1}}$ \dotfill & $\mathcal{N}(4915.61,131.51)$ & \wasponetwooneGPsigmaNIRPSone \\
$\mathrm{\rho_{GP,NIRPS1}}$ \dotfill & $\mathcal{N}(0.1011,0.0017)$ & \wasponetwooneGPrhoNIRPSone \\
$\mathrm{\sigma_{GP,NIRPS2}}$ \dotfill & $\mathcal{N}(4987.01,21.98)$ & \wasponetwooneGPsigmaNIRPStwo \\
$\mathrm{\rho_{GP,NIRPS2}}$ \dotfill & $\mathcal{N}(0.10017,0.00029)$ & \wasponetwooneGPrhoNIRPStwo \\
$\mathrm{e_{b}}$ \dotfill & $\mathrm{fixed}$ & \wasponetwooneeccpone \\
$\mathrm{\omega_{b}}$ \dotfill [$\degr$]& $\mathrm{fixed}$ & \wasponetwooneomegapone \\
\hline
$\mathrm{a_{b}}$ \dotfill [au] & $-$ & \wasponetwooneab \\ 
$\mathrm{i_{b}}$ \dotfill [$\degr$] & $-$ & \wasponetwooneincb \\ 
$\mathrm{T_{14,b}}$ \dotfill [h] & $-$ & \wasponetwooneTdurb \\ 
$\mathrm{M_{b}}$ \dotfill [$\mathrm{M_{J}}$] & $-$ & \wasponetwoonemassjb \\ 
$\mathrm{R_{b}}$ \dotfill [$\mathrm{R_{J}}$] & $-$ & \wasponetwooneradjb \\ 
$\mathrm{\rho_{b}}$ \dotfill [$\mathrm{g \, cm^{-3}}$] & $-$ & \wasponetwoonerhoplb \\ 
$\mathrm{T_{eq,b}}$ \dotfill [K] & $-$ & \wasponetwooneteqb \\ 
$\mathrm{S_{b}}$ \dotfill [$\mathrm{S_{e}}$] & $-$ & \wasponetwooneSb \\ 
\hline 
\hline
\end{tabular} 
}
\label{tab:juliet-WASP121}
\end{center} 
\begin{minipage}{\linewidth}
    %\vspace{0.1cm}
    \small Notes: Top: Fitted parameters. Bottom: derived orbital parameters and physical parameters.
    \end{minipage}
\end{table} 

%--------------------------------------------------------------------

\section{Data processing for atmospheric analyses}
\label{section:data_processing}
We carry out cross-correlation and atmospheric retrieval analyses on the combined HARPS, NIRPS and CRIRES+ transit time-series. 
% The data must be cleaned and detrended to get rid of systematic effects, and telluric and stellar contributions in order to isolate the planetary signal. 
In order to isolate the planetary signal, the data must first be detrended in order to remove telluric and stellar contributions that dominate the observed spectra. 
This involves the steps described in~\cite{pelletier_where_2021, pelletier_vanadium_2023}, and recapitulated as follow: 
We apply a $\sigma$-clipping to the flux variation and all outliers at more than 6$\sigma$ are flagged and removed. We also discard those bad exposures affected by residual systematics.
Heavily telluric contaminated regions are masked out (e.g., 0.6868-0.694 $\mu$m, 1.345-1.4465 $\mu$m, 1.7995-1.96 $\mu$m). Near-infrared to infrared spectra are notoriously polluted by Earth's atmospheric emission (i.e. OH) and absorption lines (i.e. H$_2$O, CH$_4$, CO$_2$, O$_2$). An additional mask-clipping is applied to data below 0.5 (considering the spectrum as normalised to one) to mask deep telluric spectral lines outside those highly telluric contaminated regions already discarded.
All spectra are adjusted to the same continuum level according to a double filtering approach~\citep{gibson_detection_2020}: a first box-filter of width 51 pixels is applied to each exposure, smoothed by a Gaussian convolution with a standard deviation of 100 pixels applied to each individual order. This accounts for continuum variations or blaze variations along the observed time-series.

A second order polynomial is used to fit the built out-of-transit median spectrum for one night (done night by night) and then divided from each single exposure to remove first-order stellar or telluric residual lines.
The Principal Component Analysis (PCA) routine is employed to remove any telluric lines and stellar residuals in the data and reconstruct the observed spectral temporal series. After testing, we decide to remove 5 principal components to best compromise on the cleanliness of the data while preserving the planetary signal. Finally, all column pixels having a standard deviation greater than 5$\sigma$ that of the spectral order are masked out. 
Results of the detrending steps for NIRPS, HARPS, and CRIRES+ transit time-series are shown in Figure~\ref{fig:appendix_detrending_procedure} in Appendix~\ref{section:appendix_deterending_steps}.

\subsection{Cross-correlation analysis}
\begin{figure}
   \centering
   \includegraphics[width=1.02\columnwidth]{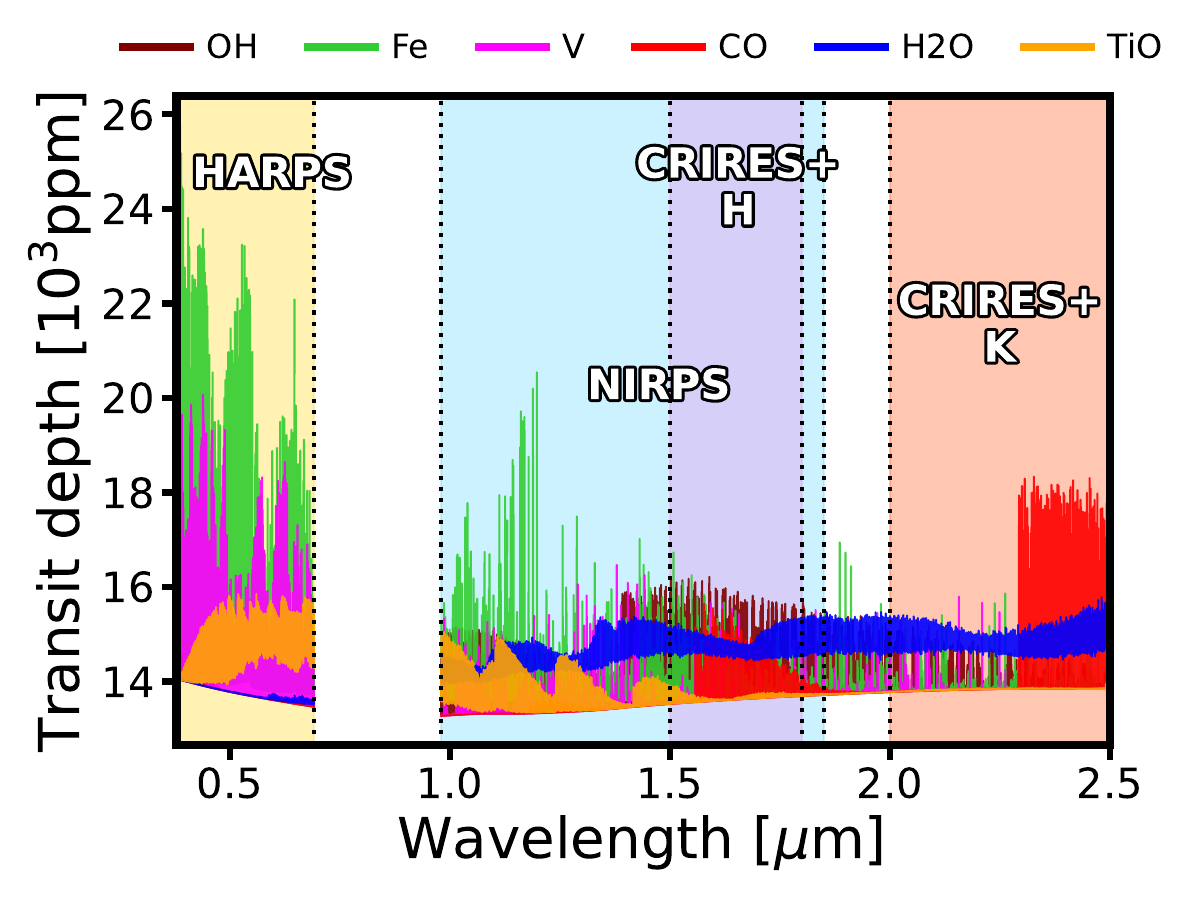}
   \caption{Atmospheric models (transit depth as function of wavelength) for OH (brown), Fe (green), V (magenta), CO (red), H$_2$O (blue), and TiO (orange) generated with the \texttt{SCARLET} code. The shaded regions highlight the spectral coverage of each instrument: the range of wavelength of HARPS is in yellow, NIRPS region in cyan, CRIRES+ $H$-band and $K$-band in purple and orange, respectively.}
   \label{fig:SCARLETmodels}
\end{figure}
As described by~\cite{vaulato+2025_hydride}, we use \texttt{SCARLET}~\citep{benneke_atmospheric_2012, benneke_how_2013, benneke_strict_2015, benneke_exoplanet_2019, pelletier_where_2021} to generate synthetic transmission spectra of \hbox{WASP-121b}.~\texttt{SCARLET} models use (i) H$_2-$H$_2$ and H$_2-$He collision-induced opacities~\citep{borysow_collision-induced_2002}, (ii)~molecular cross-sections for H$_2$O~\citep{polyansky_exomol_2018}, OH~\citep{rothman_hitemp_2010}, CO~\citep{rothman_hitemp_2010, li_rovibrational_2015}, and TiO~\citep{mckemmish_exomol_2019}, (iii)~atomic cross-sections for Fe, Ti, V, Mn, Mg, Ca, Cr, Ni, Na, Y, Ba, Sc, C from the VALD database~\citep{ryabchikova_major_2015}, and (iv)~continuum cross-sections for H$^-$ (both bound-free and free-free regimes) from~\cite{gray_observation_2021}. Molecular and atomic cross-sections are computed using \texttt{HELIOS-K}~\citep{grimm_helios-k_2015, grimm_helios-k_2021} and chemical equilibrium abundances using \texttt{FastChem}~\citep{stock_fastchem_2018, stock_fastchem_2022}. For our \texttt{SCARLET} models (Figure~\ref{fig:SCARLETmodels}), we set up a uniform and isothermal temperature structure%~\citep[the measured equilibrium temperature is $\mathrm{T\sim2\,358~K}$,][]{delrez_wasp-121_2016}%\citep[the measured dayside temperature is $\mathrm{T\sim3\,000~K}$][]{mikal-evans_diurnal_2022}
, and a cloud-free atmosphere, which is a reasonable assumption considering the high temperatures of ultra-hot gas giant atmospheres, particularly on daysides expected to be cloud-free~\citep{sing_continuum_2016, helling_cloud_2021}. %being \hbox{WASP-121b} in the regime of ultra-hot gas giants. 

The fiducial atmospheric model, initially at resolution $R=250\,000$, is convolved with a rotational kernel (accounting for the planet's synchronous rotation), %synchronous rotational kernel 
coupled with a Gaussian one-dimensional kernel to match the instrumental resolution ($R\sim80\,000$ for NIRPS, $R\sim120\,000$ for HARPS, and $R\sim86\,000-92\,000$ for CRIRES+ for the used slit width~$=0.2$"). For the cross-correlation step, we generate \hbox{WASP-121b}'s transmission model spectrum assuming a 1~$\times$~solar metallicity atmosphere at temperature of 3\,000~K. 

Once the atmospheric model has been tailored, we calculate the cross-correlation function~\citep[CCF;][]{snellen_orbital_2010} between the template spectrum and the cleaned data by Doppler shifting the model from velocity $\mathrm{v}=-400$ to $+400~\mathrm{km \, s^{-1}}$ with steps of 2~$\mathrm{km \, s^{-1}}$:
\begin{equation}
   \label{eq:CCF_function}
      \mathcal{CCF}(\mathrm{v, t}) = \sum_{i=0}^{N} \mathfrak{r}_{i}(\lambda, \mathrm{t}) T_{i}(\mathrm{v}),
   \end{equation}
where $\mathfrak{r}_{i}(\lambda, t)$ is the time series of $N$ transmission spectra as function of time $t$ and wavelength $\lambda$, namely each pixel in the transmission spectrum at a given time $t$, and $T_{i}(v)$ are the computed model spectra Doppler shifted to a velocity $\mathrm{v}$. We refer the reader to Section 4.3 in~\cite{vaulato+2025_hydride} for further details. 
The computed CCFs as in Equation~\ref{eq:CCF_function} are displayed as a function of orbital phase in two-dimensional cross-correlation trail maps for each tested chemical species. %Among the computed CCFs, we discard those cross-correlation values affected by the Doppler shadow in the velocity space~\citep{Casasayas-Barris+2022}. The stellar contamination impacts particularly HARPS optical wavelengths. On account of this, we apply a $10~\mathrm{km \, s^{-1}}$ wide mask centred at $+$38~$\mathrm{km \, s^{-1}}$~\citep[at the expected systemic velocity,][]{delrez_wasp-121_2016}.

To account for stellar contamination from the Doppler shadow, which, in our case, primarily affects HARPS optical wavelengths, we apply a $10~\mathrm{km~s^{-1}}$ wide mask centred at approximately $+35~\mathrm{km~s^{-1}}$—close to the expected systemic velocity~\citep[$+38.35\pm0.02~\mathrm{km~s^{-1}}$;][]{delrez_wasp-121_2016}~(Figure~\ref{fig:appendix-DopplerShadow}). This masking removes cross-correlation values affected by the Doppler shadow in velocity space~\citep{Casasayas-Barris+2022}.
The CCF values are then phase-folded for different orbital configurations, namely as varying the systemic velocity ($\mathrm{V_{sys}}$) from $-$100 to $+$100~$\mathrm{km \, s^{-1}}$ (1~$\mathrm{km \, s^{-1}}$ step), and the orbital velocity ($\mathrm{K_p}$) from 0 to $+$300~$\mathrm{km \, s^{-1}}$ (1~$\mathrm{km \, s^{-1}}$ step). The individual integrations are then ``stacked'' across the entire transit window to enhance a strong enough planet signal which can be interpreted as a time average being summed over the orbital phases. The results are velocity-velocity diagrams called $\mathrm{K_p}-\mathrm{V_{sys}}$ maps~\citep{brogi_signature_2012}. The maximum signal-to-noise ratio (SNR) is measured when CCFs are co-added along the Keplerian planetary trail in the velocity space. In this work, the cross-correlation analysis is run for each instrument separately. %It is worth mentioning that detections via cross-correlation analysis are not the main scope of the paper, given the recently published wealth of species detected by~\cite{prinoth+2025_W121b_4UT} leveraging the power of ESPRESSO in 4-UT mode (see their Figures~1 and 2).

\subsection{Atmospheric retrieval recipe}
%----------------------------------------
We perform a retrieval analysis to constrain a range of key atmospheric parameters, as summarized in Table~\ref{Table:retrieval_results best-fit}. These include the volume mixing ratios (VMRs) of several atomic and molecular species—specifically, $\log_{10}\mathrm{H_2O}$, $\log_{10}\mathrm{CO}$, $\log_{10}\mathrm{OH}$, $\log_{10}\mathrm{Fe}$, $\log_{10}\mathrm{V}$, and $\log_{10}\mathrm{H^-}$. The VMR of a species $\mathcal{X}$ is defined as the logarithm (base 10) of its volume mixing ratio:
\begin{align}
\label{equation:VMR}
\log_{10}\mathrm{VMR}_\mathcal{X} := \log_{10}\mathcal{X}.
\end{align}
In addition, we retrieve the electron density ($\log_{10}\mathrm{e}^-$), atmospheric temperature ($T$), orbital velocity semi-amplitude of the planet ($K_p$), systemic velocity ($V_{\mathrm{sys}}$), and the rotational broadening full width at half maximum (FWHM).
To infer the posterior probability distributions, we use a high-resolution retrieval prescription~\citep{brogi_retrieving_2019, gibson_detection_2020, gibson_relative_2022} which utilises \texttt{SCARLET} to generate the transmission spectra as detailed in~\cite{pelletier_where_2021, pelletier_vanadium_2023, pelletier_crires_2024, bazinet_subsolar_2024, vaulato+2025_hydride}. The Bayesian framework uses the likelihood approach from~\cite{gibson_detection_2020} and the \texttt{emcee} code as sampler~\citep{foreman-mackey_emcee_2013}. We run the atmospheric retrieval on the HARPS, NIRPS and CRIRES+ transits simultaneously. We also include the TESS photometric point from~\cite{bourrier_hot_2020} to ensure all sampled models match the true planetary radius. 
% For the present science case, we opt for an atmospheric ``free'' retrieval approach rather than assuming chemical equilibrium, allowing metallicities of volatile and refractory elements to be fitted independently rather than enforcing equilibrium chemistry. The free retrieval assumes the chemistry to be well-mixed (constant-with-altitude, hence uniform in pressure), the elemental abundances are individually fitted, and the chemistry predictions are not baked in the retrieval process. We assume the model to be cloud-free.
We opt for an atmospheric ``free'' retrieval approach, allowing abundances of individual species to be fitted independently as free parameters, rather than enforcing any assumptions about the atmospheric chemistry (e.g., equilibrium chemistry). The free retrieval takes input priors on the abundance profiles to be well-mixed (constant-with-altitude, hence uniform in pressure). We assume a uniform prior distribution, $\mathcal{U}$(-12, 0), for all absolute abundance. Assuming the high temperature of \hbox{WASP-121b} prevents the formation of optically-thick clouds, the continuum level is set by the H$^-$ continuum, which is parametrised with the abundances of H$^-$ and e$^-$.
We set up the free retrieval to fit for 
%The retrieval fits for 
%individual absolute abundances of H$_2$O and OH as representative of thermal dissociation processes, and CO as the main carbon-bearing species in ultra-hot Jupiters atmosphere. 
abundances of H$_2$O and OH as representative of thermal dissociation processes, according to the following chemical reactions~\citep{parmentier_thermal_2018}:
\begin{align*}
    &\ce{\mathrm{H_2O \rightleftarrows 2H + O}}~,\\
    &\ce{\mathrm{O + H2 \rightarrow H + OH~.}}
\end{align*}
Moreover, we retrieve the abundance of CO as the main carbon-bearing species in ultra-hot Jupiters atmosphere, TiO acting as primary opacity source in optical wavelength, and the abundance of refractory species detected via cross-correlation (e.g. Fe and V). We let the retrieval free to fit for the abundance of $\mathrm{H}^-$ and the electron density, used to infer the continuum opacity given by H$^-$ bound-free and free-free absorption transitions~\citep{arcangeli_h-_2018, parmentier_thermal_2018, vaulato+2025_hydride}.
%We let the retrieval free to fit for log$\alpha_{\mathrm{H}^-}$ and log$\alpha_{\mathrm{e}^-}$, multiplicative factors of the hydride and e$^-$ abundance profiles~\citep{vaulato+2025_hydride} used to calculate the continuum opacity given by H$^-$ bound-free and free-free absorption transitions~\citep{Arcangeli+2018, Parmentier+2018}.

Other than abundance parameters, the retrieval fits for the atmospheric temperature, the orbital velocities $\mathrm{K_p}$ and $\mathrm{V_{sys}}$, and a Gaussian rotational broadening parameter to account for the observed signal appearing more ``blurred'' in the $\mathrm{K_p}-\mathrm{V_{sys}}$ space than what would be expected from the instrumental resolution alone. We assume that an isothermal profile is likely a valid approximation for the average thermal structure probed in transmission, based on~\cite{Gandhi+2023} (see their Figure~11), and~\cite{maguire_high-resolution_2023} (see their Figure~3, retrieved temperature-pressure profile). 
% Despite most ultra-hot Jupiters show thermal inversion, that is not the case of \hbox{WASP-121b} at the pressure probed by our observations (sub-to-millibars) as retrieved by~\cite{Gandhi+2023} (see their Figure~11), and~\cite{maguire_high-resolution_2023} (see their Figure~3, retrieved temperature-pressure profile). 

%VV Free retrieval and why 
%VV free parameters of the fit 
%VV kp, vsys fitted not fixed and why 
%VV broadening param and why 
%VV temperature structure assumption and why (gandhi+2023)

%--------------------------------------------------------------------
\section{Results}
\label{section:results}

\subsection{Detections of cross-correlation functions}
\label{section:cross-correlation_results}
\begin{figure*}
   \centering
  \includegraphics[width=\textwidth]{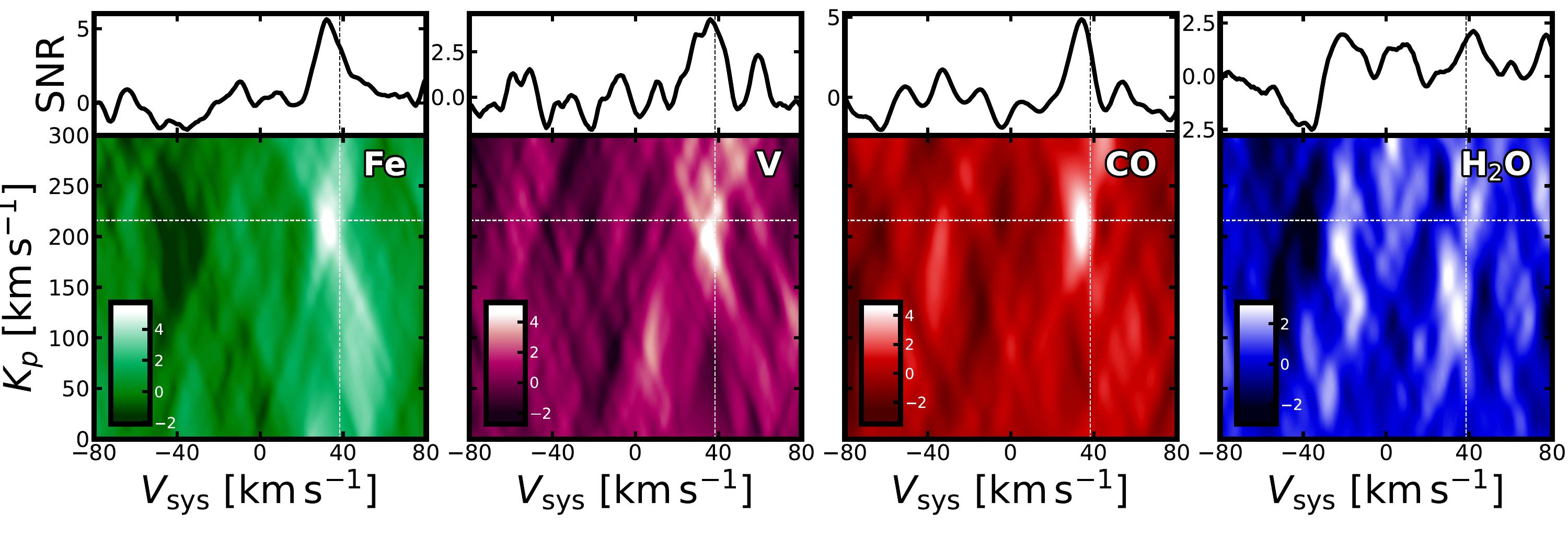}
   \caption{Detections of CCFs for Fe~(SNR$=$5.8), CO~(SNR$=$5.0), and V~(SNR$=$4.7), and non-detection of H$_2$O (SNR$\lesssim$2.0) in the transmission spectrum of \hbox{WASP-121b}. Each squared panel shows the two-dimensional cross-correlation map in the velocity-velocity space ($\mathrm{K_p-V_{sys}}$). The white dashed lines indicate the expected position of the planetary signal within the velocity-velocity space, namely the expected orbital~\citep[$\mathrm{\sim216~km~s^{-1}}$,][]{hoeijmakers_mantis_2024} and systemic ~\citep[$\mathrm{38.35\pm0.02~km~s^{-1}}$,][]{delrez_wasp-121_2016} velocities of \hbox{WASP-121b}. The rectangular panels on top depicts the one-dimensional cross-correlation of each species as signal-to-noise ratio versus systemic velocity. Detections of Fe and V are primarily driven by the optical HARPS observations, while the CO detection by the infrared CRIRES+ data (mostly $K$-band observations).}
   \label{fig:CCF_maps}
\end{figure*}
% VV add CCF map of H2O 
% VV blueshifted signatures in Vsys space
% data quality of CRIRES+ data

We detect the signatures of Fe (SNR$=$5.8) and V (SNR$=$4.7) by combining HARPS optical spectra, and CO (SNR$=$5.0) from CRIRES+ spectra in the $K$-band (Figure~\ref{fig:CCF_maps}). We thus confirm previous results reported in Section~\ref{sec:Introduction}. Neither NIRPS nor CRIRES+ $H$-band spectra yield a statistically significant H$_2$O signal (SNR$\lesssim$2.0) in the $\mathrm{K_p}-\mathrm{V_{sys}}$ space (see H$_2$O map in Figure~\ref{fig:CCF_maps}). However, the latter datasets contribute at constraining H$^-$, OH, and H$_2$O individual abundances in the retrieval analysis, as discussed in Sect.~\ref{section:retrieval_results}. The non-detection of refractory elements (Fe or Ti) in the NIRPS range could be due to the fewer and weaker spectral lines exhibited by these species compared to those in the optical range of HARPS (Figure~\ref{fig:SCARLETmodels}). However,~\cite{vaulato+2025_hydride} showed that in the case of NIRPS observations of the ultrahot gas giant WASP-189b, these lines are damped below $\sim 1.6~\mu\mathrm{m}$ by the H$^-$ bound-free continuum \citep[][see also Fig. 7 in~\citeauthor{lothringer_extremely_2018}~\citeyear{lothringer_extremely_2018}]{arcangeli_h-_2018, parmentier_thermal_2018}. 

The SNR maxima of the one-dimensional cross-correlation functions of all three detected signals (top panel in Fig.~\ref{fig:CCF_maps}) are slightly offset ($<10~\mathrm{km \, s^{-1}}$, more specifically the signal peaks between 31-36~$\mathrm{km \, s^{-1}}$) with respect to the reference $\mathrm{V_{sys}\sim38.35~\mathrm{km \, s^{-1}}}$~\citep{delrez_wasp-121_2016}. This could indicate the presence of winds at sub-millibar pressures (see Sect.~\ref{section:orbital_velocities}). The consistent blueshifted signals observed for the refractory elements (e.g. Fe and V) are in agreement with sub-to-antistellar circulation in atmospheric layers, as reported by~\cite{seidel+2025_w121b_dynamics_nature}~(see their Figure 2). Alternatively, systematic offsets (different instrument by instrument) or instrumental noise might be responsible for the observed blueshifted features, hence the offset might have a non-physical meaning (not related to atmospheric circulations). However, it is worth noticing that the blueshift is fairly consistent between the HARPS-driven Fe and V detections, and the CRIRES+-driven CO detection, further supporting the hypothesis of day-to-night winds.

\subsection{Retrieval results}
\label{section:retrieval_results}
\begin{table*}[!t]
    \centering
    \def\arraystretch{1.2}
    \caption{\hbox{WASP-121b} \texttt{SCARLET} atmospheric posteriors from the retrieval analysis on the HARPS, NIRPS, and CRIRES+ time series combined.}
    \begin{tabularx}{\textwidth}{>{\centering\arraybackslash}m{3cm} X >{\centering\arraybackslash}m{4cm} >{\centering\arraybackslash}m{4cm}}
        \toprule
        \toprule
        Parameters & Description & Value & Prior range \\
        \midrule
        \(\mathrm{log_{10}{Fe}}\) & Volume mixing ratio of iron& $-$6.33$^{+0.99}_{-0.40}$ & $\mathcal{U}$($-$12, 0)\\
        \(\mathrm{log_{10}{V}}\) & Volume mixing ratio of vanadium& $-$9.89$^{+0.52}_{-0.53}$ & $\mathcal{U}$($-$12, 0)\\
        \(\mathrm{log_{10}{H^-}}\) & Volume mixing ratio of hydride& $-$10.31$^{+0.71}_{-0.60}$ & $\mathcal{U}$($-$12, 0)\\
        \(\mathrm{log_{10}{e^-}}\) & Volume mixing ratio of electrons& $-$6.08$^{+3.22}_{-3.67}$ & $\mathcal{U}$($-$12, 0)\\
        \(\mathrm{log_{10}{H_2O}}\) & Volume mixing ratio of water vapour& $-$6.52$^{+0.49}_{-0.68}$ & $\mathcal{U}$($-$12, 0)\\
        \(\mathrm{log_{10}{CO}}\) & Volume mixing ratio of carbon monoxide& $-$5.29$^{+0.69}_{-0.87}$ & $\mathcal{U}$($-$12, 0)\\
        \(\mathrm{log_{10}{TiO}}\) & Volume mixing ratio titanium oxide& $-$11.18$^{+0.66}_{-0.41}$ & $\mathcal{U}$($-$12, 0) \\
        \(\mathrm{log_{10}{OH}}\) & Volume mixing ratio of hydroxil& $-$8.22$^{+1.24}_{-2.52}$ & $\mathcal{U}$($-$12, 0) \\
        T [K] & Atmospheric temperature & 2\,828$^{+691}_{-238}$ & $\mathcal{U}$(100, 6000) \\
        $\mathrm{K}_\mathrm{p}$ [km/s] & Planet velocity semi-amplitude & 202.99$^{+2.84}_{-2.92}$ & $\mathcal{U}$(156, 276) \\
        $\mathrm{V}_\mathrm{sys}$ [km/s] & Systemic velocity & $31.96^{+0.71}_{-0.69}$ & $\mathcal{U}$(18, 58) \\
        FWHM [km/s] & Rotational broadening FWHM & 6.20$^{+1.40}_{-1.67}$ & $\mathcal{U}$(0, 20) \\
        \bottomrule
        \(\log(\mathrm{H^-/Fe})\) & Relative abundance & $-$3.98$^{+0.81}_{-1.14}$ & -- \\
        \(\log(\mathrm{V/Fe})\) & Relative abundance & $-$3.56$^{+0.66}_{-1.11}$ & -- \\
        \(\log(\mathrm{CO/Fe})\) & Relative abundance & $1.04^{+0.80}_{-1.32}$ & -- \\
        \(\log(\mathrm{H_2O/CO})\) & Relative abundance & $-$1.23$^{+1.00}_{-0.97}$ & -- \\
        \bottomrule
        \bottomrule
    \end{tabularx}
     \label{Table:retrieval_results best-fit}
\end{table*}

\begin{figure*}[!h]
    \centering
    \includegraphics[width=\textwidth, trim=47 0 47 5, clip]{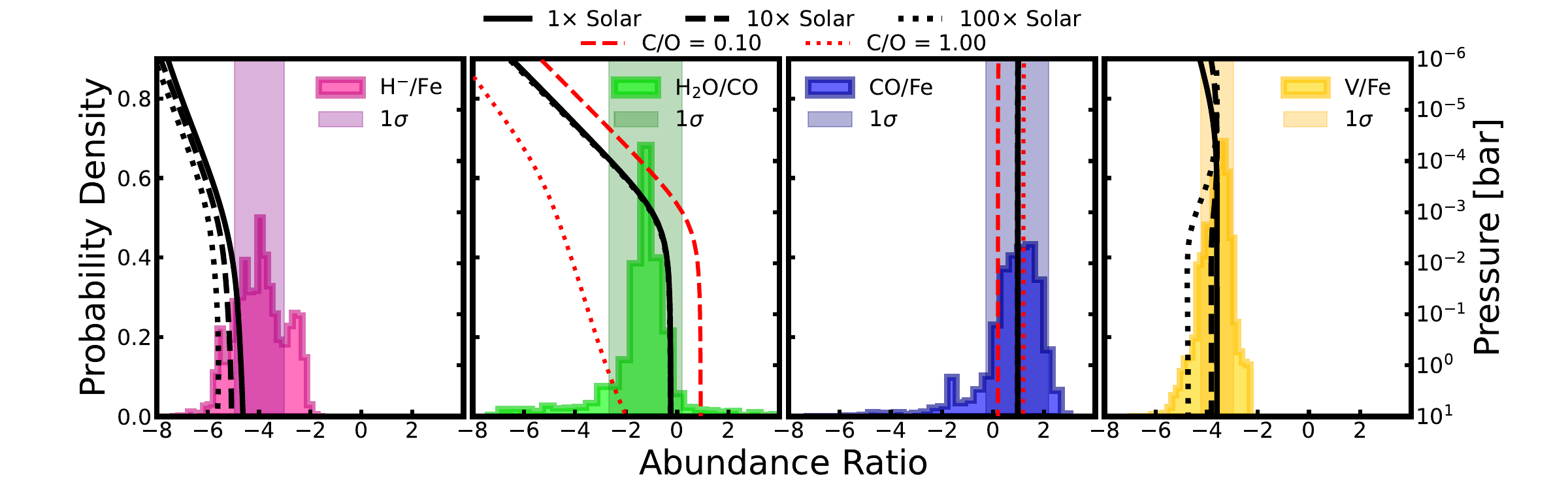}
    \caption{%\textcolor{red}{Plot updated at June 25th 2025.} 
    Key retrieved results. The four distributions represent the probability density of retrieved chemical abundance ratios (in logarithmic scale). Overplotted the \texttt{FastChem} model predictions in chemical equilibrium as varying the metallicity and the C/O ratio. \texttt{FastChem} theoretical models are displayed as function of the pressure in bar.}
    \label{fig:retrieval_results}
\end{figure*}

\begin{figure*}[!h]
    \centering
    \includegraphics[width=\textwidth]{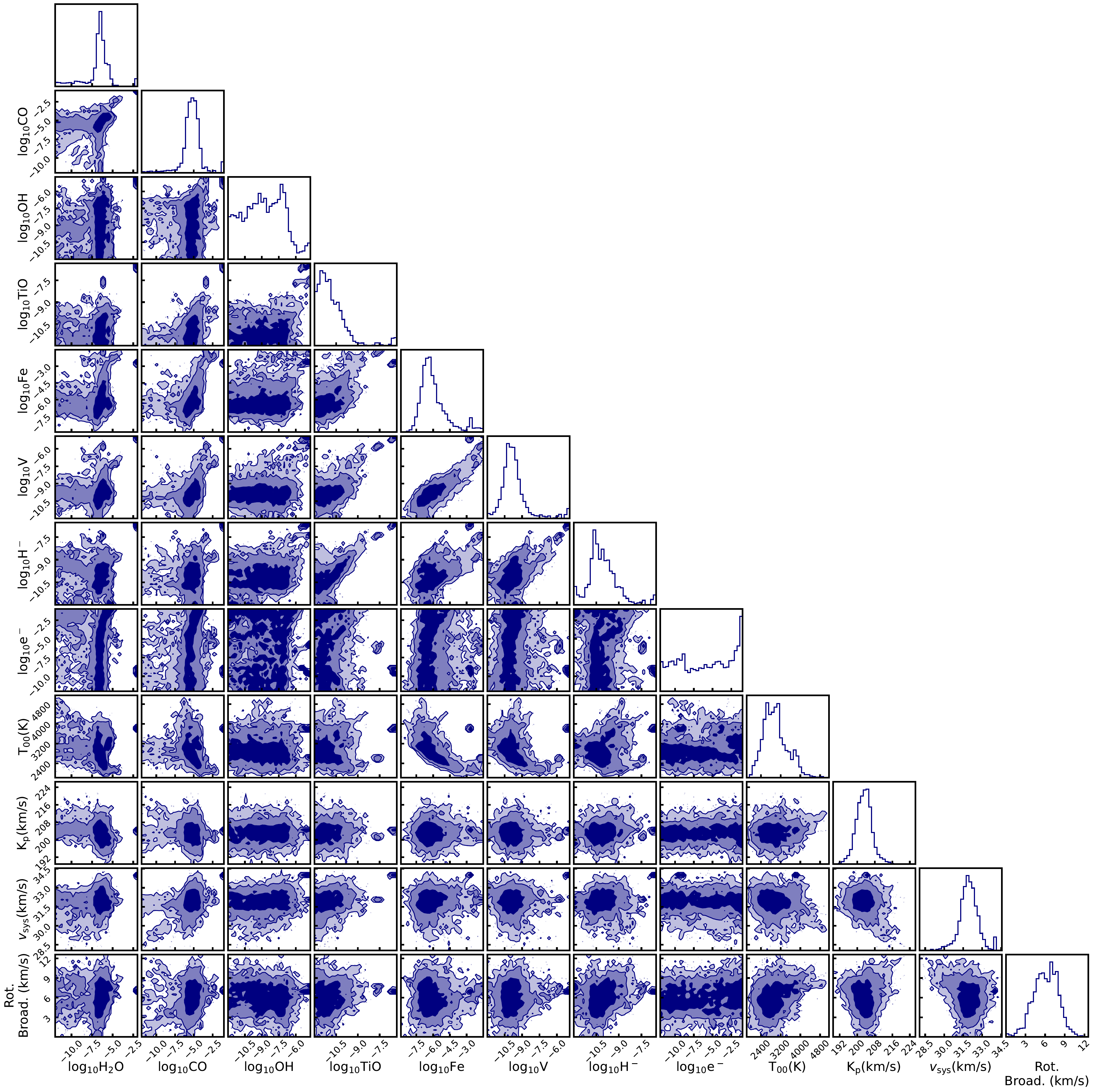}
    \caption{%\textcolor{red}{Plot updated at June 25th 2025.} 
    Retrieved constraints on the atmospheric and orbital properties from HARPS, NIRPS and CRIRES+ combined transits of \hbox{WASP-121b}. We assume a free atmospheric retrieval condition, hence a well-mixed chemistry (constant-with-altitude).}
    \label{fig:corner_plot_retrieval_results}
\end{figure*}

We summarise the key results of the free atmospheric retrieval analysis on \hbox{WASP-121b} HARPS, NIRPS, and CRIRES+ time series combined in Figure~\ref{fig:retrieval_results}, Figure~\ref{fig:corner_plot_retrieval_results}, and Table~\ref{Table:retrieval_results best-fit}. 
We retrieve the best-fit abundances for the chemical species injected in our model (i.e.\ Fe, V, CO, H$_2$O, OH, TiO, H$^-$, and e$^-$). However, retrieved absolute abundances are not informative of the underlying chemistry in a planet's atmosphere, as they significantly rely on model assumptions and, thus, are model dependent.
%Furthermore, retrieved absolute abundances rely on abundance constraints that are generally broad, with uncertainties of the order of 1~dex, and dependent on prior distributions and the assumed scattering properties. Thus, this leads to strong correlations that indicate that relative abundances of species are constrained more accurately than absolute individual abundances~\citep[][see their Section 5]{gibson_relative_2022}.
Furthermore, retrieved absolute abundances rely on abundance constraints that are generally broad, with uncertainties of the order of 1~dex, and dependent on prior distributions~\citep[][see also~\citeauthor{burningham+2017_retrieval_atmo_Ldwarfs}~\citeyear{burningham+2017_retrieval_atmo_Ldwarfs}]{gibson_relative_2022}. Namely, the spectra can be well-fitted by models with different %input%configuration
parameters (e.g.~reference pressure, continuum level, or planet radius). Inferring the volume mixing ratios of one specific absorbing gas of the atmosphere by directly measuring its absorption features is challenging because it depends on different combinations of the planet's radius and the mixing ratio assumed, yielding the same number density. For instance, two atmospheres with different absorber mixing ratios might have consistent pressure and densities leading to similar absorption spectral features~\citep[][see their Figure 4 and Section 4.1.1]{benneke_atmospheric_2012}.
Thus, this leads to strong correlations indicating that relative abundances of species are constrained more accurately than absolute abundances~\citep[][see their Section 5]{gibson_relative_2022}.
%In addition, individual abundances are challenging to interpret in the context of ongoing chemical and physical processes in the atmosphere. 
%In addition, individual abundances are difficult to interpret meaningfully without considering their ratios to other species, as chemical and physical processes in the atmosphere are best understood through relative rather than individual abundances. %Relative chemical abundances (abundance ratios between species) are rather more likely to provide insights into how each species contributes to the overall budget of volatile and refractory compounds, besides the ongoing chemical processes. 
%Inferred relative chemical abundances 
%%(abundance ratios between species) 
%are more reliable since relative line strengths are preserved even if all continuum information are lost.
% Not 100% sure here.
%Inferred relative chemical abundances 
%%(abundance ratios between species) 
%are more reliable since we remain sensitive to the spectral line strength above the continuum level.
%% Questo per spiegare perchè il key plot mostra abundance ratios di interesse attorno ai quali costruisco l'interpretazione nel contesto dei processi chimici ongoing nell'atmosfera.

To investigate the chemical composition of \hbox{WASP-121b}’s atmosphere, we calculate abundance ratios between pairs of species, $\mathcal{X}$ and $\mathcal{Y}$, using their volume mixing ratios as defined in Equation~\ref{equation:VMR}. The abundance ratio is expressed in logarithmic form as:
\begin{align}
\label{equation:relative_abundances}
 \log(\mathcal{X}/\mathcal{Y} )
&:= \log_{10} \mathcal{X} - \log_{10} \mathcal{Y} \nonumber \\
&:= \log_{10} \mathrm{VMR}_\mathcal{X} - \log_{10} \mathrm{VMR}_\mathcal{Y} \nonumber \\
&:= \log_{10} \left( \frac{\mathrm{VMR}_\mathcal{X}}{\mathrm{VMR}_\mathcal{Y}} \right) \nonumber \\
&:= \log_{10} \left( \frac{\mathcal{X}}{\mathcal{Y}} \right)
\end{align}
We constrain the relative abundances  for $\log(\mathrm{H^-/Fe})=-3.98^{+0.81}_{-1.14}$, $\log(\mathrm{H_2O/CO})=-1.23^{+1.00}_{-0.97}$, $\log(\mathrm{CO/Fe})=1.04^{+0.80}_{-1.32}$, and $\log(\mathrm{V/Fe})=-3.56^{+0.66}_{-1.11}$~(Table~\ref{Table:retrieval_results best-fit},~Figure~\ref{fig:retrieval_results}). We now compare the retrieved abundance ratios with predictions from modelled equilibrium chemistry~\citep{kitzmann_fastchem_2024}. 
Overall, the retrieved abundance ratios are consistent with a 1~$\times$~solar metallicity and a solar C/O ratio \citep[~(C/O)$_\odot=0.59\pm0.08$;][]{asplund_chemical_2021}.%, meaning that chemical processes (i.e. molecular dissociation, condensation, and vertical mixing) have not significantly altered atmospheric composition, at least at the pressure and temperature probed by our observations. 

The fraction of H$_2$O relative to CO~(Figure~\ref{fig:retrieval_results}, second panel) is 1 dex underabundant (sub-solar) than predictions from \texttt{FastChem} (but still consistent with solar within 1-$\sigma$), likely due to water being partially dissociated into OH and O~\citep{parmentier_thermal_2018}~at the high temperatures at play. This underabundance would be even more pronounced if assuming the stellar C/O rather than the solar C/O. We, indeed, retrieve a temperature $\mathrm{T=2\,828^{+691}_{-238}~K}$, assumed to be uniform across the atmosphere at the pressures probed, in agreement with~\cite{gandhi_retrieval_2023}. At this temperature, the abundance of CO for a nearly-solar C/O remains unchanged. Therefore, according to~\cite{madhusudhan_co_2012}, the $\log(\mathrm{H_2O/CO})$ ratio is expected to be close to unity for a nearly-solar C/O ratio, in agreement with our retrieved abundance ratio ($\log(\mathrm{H_2O/CO})=-$1.23$^{+1.00}_{-0.97}$). 

Intriguingly, the free retrieval constrains the abundance of H$_2$O despite the lack of a cross-correlation detection in our data (Figure~\ref{fig:CCF_maps}, forth panel). This constraint is rather driven by the retrieval preferring to add water. In fact, H$_2$O has been detected in both high- and low-resolution observations~\citep{evans_detection_2016, wardenier_phase-resolving_2024}. Therefore, a non-detection in the cross-correlation space is more likely due to the limited sensitivity of our data rather than the absence of H$_2$O in \hbox{WASP-121b} atmosphere. Notably, ~\cite{wardenier_phase-resolving_2024} reported a significant H$_2$O detection only after stacking three transits gathered with IGRINS installed on Gemini, a 8-m class telescope. The breaking up of water molecules is significant in ultra-hot gas giants, where it must be accounted for when inferring chemical abundances, especially in comparison to more thermally stable species like CO. Carbon monoxide is, indeed, a benchmark molecule to compare with~\citep{parmentier_thermal_2018, savel_diagnosing_2023}. The reason is twofold: (i)~due to its chemical robustness (thanks to the triple covalent bond), CO is less prone to thermal dissociation (via~\hbox{CO~$\rightleftarrows$~C~+~O}), thus remaining nearly constant with pressure~\citep[with the exception of known torrid exoplanets with fully dissociated atmospheres, such as KELT-9b,][]{gaudi_giant_2017, hoeijmakers_atomic_2018}; and, (ii)~CO is also poorly sensitive to model assumptions on its abundance profile, being resistant to thermal dissociation. The $\log(\mathrm{H_2O/CO})$ abundance ratio also serves as an indicator of the atmospheric region probed by our data, as well as of the dissociation and condensation of volatile compounds. Indeed, water condensates and dissociates more readily than CO for the reasons illustrated above. According to the retrieved quantity, which relies on model assumptions, and the chemistry predictions, we sample the pressure layer from sub-to-millibar ($\sim10^{-4}-10^{-3}$~bar), hence, we do not have access to the inflated upper atmosphere of \hbox{WASP-121b}, but rather we have insights to lower pressures. 

We find that the retrieved $\log(\mathrm{CO/Fe})$ ratio is consistent with the solar value (Figure~\ref{fig:retrieval_results}, third panel), implying that the planet's atmosphere likely experiences moderate-to-high temperature, not extreme enough to dissociate CO molecules and not cool enough to allow extensive Fe rainout. %~\citep[as in WASP-76b,][]{ehrenreich_nightside_2020}.
The $\log(\mathrm{CO/Fe})$ ratio can be also interpreted as a proxy of the volatile-to-refractory budget, i.e. the ice-to-rock ratio.
Moreover, the $\log(\mathrm{CO/Fe})$ ratio, being consistent with a model assuming a solar amount of carbon relative to oxygen, suggests that the atmosphere is not heavily skewed toward either a super-solar abundance of carbon or oxygen, but is instead chemically balanced. However, an atmosphere carrying a C/O ratio consistent-with-solar, likely favour a surplus of oxygen which is mostly bound in water molecules at low temperatures but which is set free by thermal dissociation of H$_2$O at high temperatures (as the case of ultra-hot gas giants) and pairs to form CO (CH$_4$+H$_2$O~$\rightleftarrows$~CO+3H$_2$) and OH (H$_2$O~$\rightleftarrows$~2H+O)~\citep{madhusudhan_co_2012}.

The retrieved consistency with a  C/O$=$solar contrasts with: (i) the carbon-enriched (volatile-enriched relative to refractory) atmosphere reported by~\cite{pelletier_crires_2024}, which shows a super-solar C/O ratio (C/O~$=0.87^{+0.04}_{-0.06}$) from a free retrieval and a super-stellar ratio (C/O~$=0.73^{+0.07}_{-0.08}$) from a chemically consistent retrieval; and (ii) the super-stellar C/O ratio (C/O~$=0.70^{+0.07}_{-0.10}$) and a moderately super-stellar refractory-to-volatile ratio ($3.83^{+3.62}_{-1.67}\times\mathrm{stellar}$) found by~\cite{smith_roasting_2024} under chemical equilibrium conditions. Despite both studies report elevated C/O ratios from dayside observations,~\cite{pelletier_crires_2024} favour a scenario enriched in ice-forming elements over rock-forming elements, suggesting the planet formed farther out in the disk beyond the ice-line, while~\cite{smith_roasting_2024} suggest that WASP-121b likely formed between the soot line and the H$_2$O snow line~\citep{lothringer_new_2021, chachan_breaking_2023}. Our finding places WASP-121b in the nearly chemically balanced regime, not markedly favouring either a volatile-rich or refractory-rich formation environment. However, it is worth noticing that our results based on transit time-series rather than dayside observations, thus our data probe a different altitude in the atmosphere of WASP-121b. 

We retrieve a $\log(\mathrm{V/Fe})$ ratio that is consistent with chemical models assuming a solar composition (Figure~\ref{fig:retrieval_results}, forth panel) indicating that the refractory balance is preserved in the atmosphere of \hbox{WASP-121b}, fully in agreement with equilibrium chemistry predictions. Our retrieved elemental ratio is also well consistent with that of~\cite{maguire_high-resolution_2023} (see their Figure~4, second panel), based on three transits observed with ESPRESSO/VLT and fully in agreement with one UVES transit used as a benchmark.

The measured $\log(\mathrm{H^-/Fe})$ ratio (Figure~\ref{fig:retrieval_results}, first panel) can be a tracer of both the opacity contribution from H$^-$ in the~(near-)infrared~regime, and the ionisation processes involving the electron density associated with the ionisation of metals, which are the primary source of free electrons~\citep{bell_free-free_1987, john_continuous_1988}. %H$^-$ is known to be a non-negligible continuum, potentially muting planetary absorption features in ultra-hot gas giants~\citep{vaulato+2025_hydride} if found overabundant than what predicted by equilibrium chemistry.

The retrieval fails to constrain the abundance of free electrons (Figure~\ref{fig:corner_plot_retrieval_results}), likely due to either the spectral range of our observations does not extend to wavelengths sensitive to the free-free transitions to form H$^-$, or the inefficiency of metal ionisation within the pressure and temperature regime probed by the data. 
H$^-$ bound-free absorption can dominate the spectrum below $\sim1.6~\mu$m, masking spectral features from species such as H$_2$O and TiO. %This may explain why water molecules are constrained by the retrieval but not detected in cross-correlation (Figure~\ref{fig:CCF_maps}). 
Similarly, the retrieval sets an upper limit for TiO, despite also non-detected in cross-correlation. Nevertheless, caution is warranted, as the line list used for TiO may lack the accuracy required to draw firm conclusions about the presence of TiO in the atmosphere of \hbox{WASP-121b}~\citep{prinoth+2025_W121b_4UT}. However,~\cite{prinoth+2025_W121b_4UT} reported a significant detection of Ti (albeit at a weaker strength then expected) in high signal-to-noise optical transits, suggesting that TiO may also be present.

In ultra-hot gas giants, ionisation involves both thermal processes and photoionisation~\citep{kitzmann_peculiar_2018, fossati_data-driven_2020}. However, the effects of the latter is not considered in the equilibrium chemistry models. Hence, the modelled $\log(\mathrm{H^-/Fe})$ profile primarily reflects thermally driven ionisation processes occurring in \hbox{WASP-121b} at (sub-)millibar pressures. %We surmise that photoionisation enters the picture higher up in the atmosphere (at lower pressures) not probed by our observations. 
In this context, the measured $\log(\mathrm{H^-/Fe})$ is consistent with \texttt{FastChem} models of a solar composition, and function as a thermometer in thermal equilibrium atmospheres according to the Saha ionisation equation~\citep{saha_liii_1920}. Even though the retrieved elemental hydride-to-Fe ratio is consistent with model predictions within 1$\sigma$, the median of the Gaussian distribution peaks at marginally super-solar values, hinting at either an elevated abundance of hydride continuum opacity similar to what found in WASP-189b~\citep{vaulato+2025_hydride} but likely to compensate for missing opacity contributions in the model assumption, or a depleted Fe content due to condensation. The atmosphere of WASP-121b boasts a wealth of detected chemical species (see Section~\ref{sec:Introduction} for a complete literature overview) which we decided not to include in our atmospheric retrieval calculation and model. The main reason behind our choice is computational limitations. Therefore, H$^-$ is supposedly overcompensating for other missing opacity sources from species not included in our model. Indeed, the continuum should be handled mainly by the hydride and so by the density of free-electrons (e-), which have proven to be the principal continuum absorbers, particularly in ultra-hot gas giants~\citep{parmentier_thermal_2018, arcangeli_h-_2018, vaulato+2025_hydride}. We also surmise that photoionisation mechanisms might become relevant since the $\log(\mathrm{H^-/Fe})$ ratio peaks at about 10$^{-4}$, higher than predicted by solar chemical equilibrium models (10$^{-5}$). Nevertheless, the retrieved $\log(\mathrm{H^-/Fe})$ abundance ratio results consistent with solar models within 1 dex.

%We also surmise that photoionisation mechanisms becomes relevant since the $\log(\mathrm{H^-/Fe})$ ratio is a bit higher than predicted by equilibrium chemistry. %Owing to a poorly constrained electron density, we cannot discriminate between the two scenarios. %not sure here...

\subsection{Orbital velocities}
\label{section:orbital_velocities}
%% DeltaKp = KpRetrieved - KpOrbital discussion.
%% Rotation broadening parameter
%% Wardenier+2024
%% Torres+2010
%% Sing+2025

We retrieve a systemic velocity of $\mathrm{V_{\rm sys} = 31.96^{+0.71}_{-0.69}~km~s^{-1}}.$%, consistent with previous works: $\mathrm{V_{\rm sys} \sim 38.3~km~s^{-1}}$\citep{bourrier_hot_2020}, $\mathrm{V_{\rm sys} = 38.198 \pm 0.002~km~s^{-1}}$\citep{borsa_atmospheric_2021}, and $\mathrm{V{\rm sys} = 38.64 \pm 0.06~km~s^{-1}}$~\citep{seidel_detection_2023}.
We also constrain an orbital velocity of $\mathrm{K_p = 202.99^{+2.84}_{-2.92}~km~s^{-1}}$. 
%We also constrain an orbital velocity of $\mathrm{K_p = xx \pm yy~km~s^{-1}}$, in agreement with results from the literature: $\mathrm{K_p = 216.51^{+0.28}{-0.33}~km~s^{-1}}$ (hybrid free), $\mathrm{K_p = 216.49^{+0.29}{-0.30}~km~s^{-1}}$ (chemical equilibrium)~\citep{pelletier_crires_2024}, $\mathrm{K_p \sim 217.65km~s^{-1}}$~\citep{borsa_atmospheric_2021}, $\mathrm{K_p = 211.8 \pm 7.6~km~s^{-1}}$\citep{prinoth+2025_W121b_4UT}, and $\mathrm{K_p = 213.61 \pm 0.88~km~s^{-1}}$ (dataset T1), $\mathrm{K_p = 208.80 \pm 1.65~km~s^{-1}}$ (dataset T2), and $\mathrm{K_p = 197.67 \pm 4.24~km~s^{-1}}$ (dataset T3)~\citep[][see their Table A.6]{maguire_high-resolution_2023}. 

\begin{figure}
    \includegraphics[width=1.02\columnwidth]{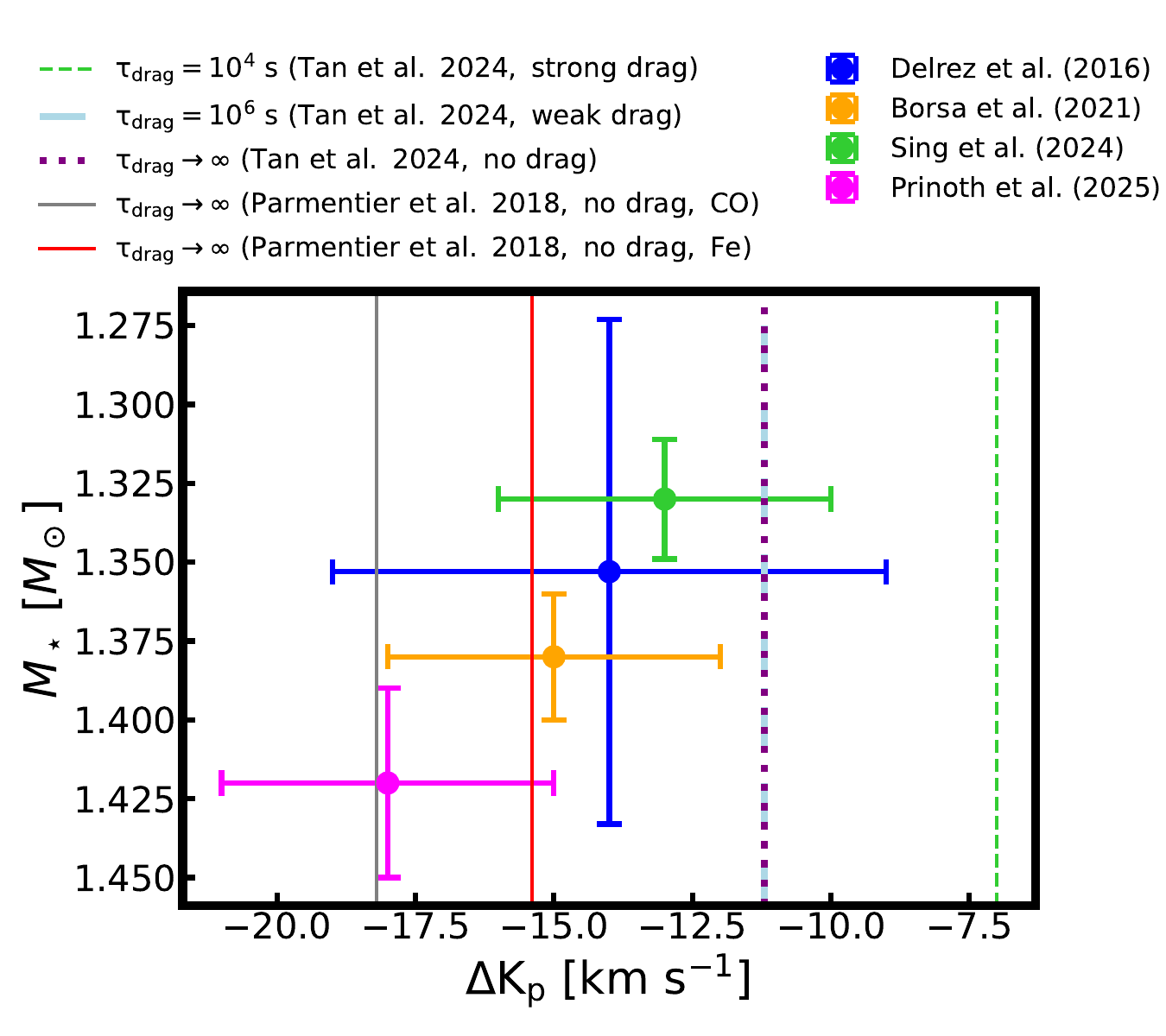}
    \caption{$\Delta\mathrm{K_p}$ as function of different stellar masses. This plot serves to visualize the Table~\ref{tab:delta_kp_values} in comparison with global circulation model predictions. Each data point with uncertainties correspond to the computed $\Delta\mathrm{K_p}$ as function of the stellar mass (reported in Table~\ref{tab:delta_kp_values}) as derived by~\cite{delrez_wasp-121_2016} ($\textrm{M}_\star=1.35\pm0.08~\mathrm{M_\odot}$),~\cite{borsa_atmospheric_2021} ($\textrm{M}_\star=1.38\pm0.02~\mathrm{M_\odot}$),~\cite{sing_absolute_2024} ($\textrm{M}_\star=1.33\pm0.02~\mathrm{M_\odot}$),~\cite{prinoth+2025_W121b_4UT} ($\textrm{M}_\star=1.42\pm0.03~\mathrm{M_\odot}$), respectively. Uncertainties are calculated via error propagation. Vertical lines mark the GCMs predictions, namely models with zero, weak and strong drag by~\cite{tan_modelling_2024}, and drag-free models by~\cite{parmentier_thermal_2018}.}
    \label{fig:deltaKp}    
\end{figure}

Leveraging the largest radial velocity dataset of \hbox{WASP-121b} to date, we update the orbital parameters, deriving the orbital velocity $\mathrm{K_p}$ from the stellar reflex motion obtained via the \texttt{juliet} fitting routine, as detailed in Section~\ref{section:RV}. By comparing our derived orbital velocities from both the RV analysis and the atmospheric retrieval routine, we calculate the velocity offset, $\Delta\mathrm{K_p}$\footnote{$\Delta\mathrm{K_p=K_p^{retrieved}-K_p^{orbital}}$}~\citep{wardenier_2025_w76b_pretransit_posteclipse}, which reflects the rate of change of the Doppler shift of the planetary signal across the orbital phases from (0,0)~$\mathrm{km~s^{-1}}$ in the planet rest frame. By inspecting the $\mathrm{K_p-V_{sys}}$ maps, the planetary signal appears overall diffused (not a point-like source), likely due to dynamics and 3D effects, which are modelled as one free rotational broadening parameter in the atmospheric retrieval and retrieved to be $6.20^{+1.40}_{-1.67~}~\mathrm{km~s^{-1}}$, consistent with the expectations. Transmission data are generally more sensitive to atmospheric dynamics, hence, we could expect a rotational broadening parameter that slightly differs from the planet rotation rate\footnote{$v_\mathrm{eq.~rot.}$ = $\omega~\times~R_\mathrm{p}=(2\pi/P)~\times~R_\mathrm{p}\simeq6.86~\mathrm{km~s^{-1}}$, where $\omega=2\pi/P$ is the angular velocity, $P$ is WASP-121b orbital period estimated with the \texttt{juliet} RV fitting carried out in this work (see Table~\ref{tab:juliet-WASP121}), and $R_\mathrm{p}$ is the planet radius.} The retrieved rotational broadening parameter is not significantly under or over-broadened, thus ruling out potential resolution biases.
In case of Fe, V and CO, the signal-to-noise ratio peaks at both lower $\mathrm{K_p}$ and lower $\mathrm{V_{sys}}$ values (see Figure~\ref{fig:CCF_maps}) than the reference ones, resulting in blueshifted signals along the radial velocity axis and a change in slope (i.e. acceleration) along the the orbital velocity axis. %The observed blueshift is in agreement with~\cite{wardenier_phase-resolving_2024}, and support a likely drag-free model. %Contrastingly, the CO detection also pops out at blueshifted $\mathrm{V_{sys}}$, aligned with the refractory species, but peaks at an interestingly higher $\mathrm{K_p}$ compared to the reference orbital velocity~\citep[$\sim216~\mathrm{km~s^{-1}}$,][]{hoeijmakers_mantis_2024}. %According to~\cite{wardenier_modelling_2023}, the observed blueshift at ingress is attributed to the leading, morning-side limb, where day-to-night winds dominate, while redshifted components are associated with planetary rotation.It is worth noting, however, that the observed $\mathrm{K_p}$ offset remains within the uncertainty of the reference value. %Not sure here
These results can be contextualized within the framework proposed by~\cite{wardenier_modelling_2023, wardenier_phase-resolving_2024} who present detailed three-dimensional global circulation models for \hbox{WASP-121b}, and simulate the resulting high-resolution transmission spectra
across different dynamical regimes. We consider four GCMs (see Figure~\ref{fig:deltaKp}): the first is a drag-free ($\mathrm{\tau_{drag}\rightarrow\infty~}$) model illustrated in~\cite{parmentier_thermal_2018}, while the other three models are presented in~\cite{tan_modelling_2024} including one drag-free ($\mathrm{\tau_{drag}\rightarrow\infty~}$) model, one weak-drag model ($\mathrm{\tau_{drag}=10^{6}~s}$), and one with a strong-drag prediction ($\mathrm{\tau_{drag}=10^{4}~s}$), where $\tau$ is the uniform drag timescale. The drag-timescale variable represents the time needed for a parcel of air to lose a substantial percentage of its kinetic energy. It describes the wind speed, thus the efficiency of heat redistribution, and the Doppler shifts observed in the spectra. %The zero-drag model accounts for a super-rotating equatorial jet, while jet formation is not considered in the other models, rather the day-to-night flow is at the base of the wind profile. However, the day-to-night contrast is preserved in all models being the planet tidally locked.
The zero-drag model accounts for a super-rotating equatorial jet~\citep{wardenier_decomposing_2021}, while jet formation is suppressed in the drag-models. Indeed, in both weak- and strong-drag models, the wind profile is mainly dominated by day-to-night flows. Moreover, the day-to-night contrast is preserved in all models being the planet tidally locked.
Therefore, these global circulation models demonstrate that the presence, or absence, of atmospheric drag strongly modulates the Doppler shift of the planetary signal. In particular, drag-free models naturally produce a symmetric latitudinal temperature structure and high longitudinal temperature gradients at the terminator (strong day-to-night temperature variations), resulting in measurable Doppler shifts in the planetary trail due to winds and rotation. Indeed, the changing and evolving structure of the planet's atmosphere as viewed at different phases results in a variation of the position of the planet's spectral lines, eventually resulting in a measurable shift in the $\mathrm{K_p}$. This is consistent with our detection of a negative $\mathrm{\Delta K_p}$ and $\mathrm{\Delta V_{sys}}$ for Fe, V and CO. Indeed, GCMs predict a decrease in the observed $\mathrm{K_p}$, specifically producing a negative $\Delta\mathrm{K_p}$, regardless of the drag.%, which is predicted in drag-free regimes where the day-night contrast is preserved and fast, eastward winds enhance the effective line-of-sight velocities at ingress and egress.

%Moreover,~\cite{wardenier_phase-resolving_2024} show that species such as Fe and CO probe different pressure levels and respond differently to dynamical conditions. %While refractory species (e.g., Fe, V) trace millibar regions where rotation and jets dominate, CO can probe slightly higher altitudes and trace more complex wind structures. 
%Our observation of the CO signal being offset in both $\mathrm{V_{sys}}$ and $\mathrm{K_p}$ — yet consistent with the signature of the refractory species in $\mathrm{V_{sys}}$ — further supports either a drag-free or a weak-drag scenario with H$_2$ dissociation/recombination, particularly if multiple layers contribute to the observed spectral features.

However, as highlighted in Table~\ref{tab:delta_kp_values} and Figure~\ref{fig:deltaKp}, the calculation of $\Delta\mathrm{K_p}$ is sensitive to the assumed stellar mass, which feeds directly into the derived orbital velocity. Differences in $\mathrm{M_\star}$ lead to non-negligible changes in the interpretation of the Doppler shift and, therefore, in how well our observations align with specific GCM predictions. This emphasizes the importance of using consistent stellar parameters when comparing observationally derived kinematic shifts to model outputs. 

WASP-121 is known as an hot, F6-type star, with a moderate chromospheric stellar activity level~\citep[$\mathrm{logR'_{HK}\simeq-4.8}$~measured by][]{borsa_atmospheric_2021}.~\cite{bourrier_hot_2020} also reported on magnetically active regions at the surface of WASP-121, visible in the periodogram of the RV residuals (see their Figure 3). Moreover, the moderate-to-fast rotation of WASP-121~\citep[$\mathrm{v~sin(i)=11.90\pm0.31~km~s^{-1}}$,][]{borsa_atmospheric_2021} broadens the few stellar spectral lines typical of A- and F-type stars, blending lines that are usually isolated and more ``peaky" in cold and slow rotating stars. As a result, deriving the chemical composition of these stars (e.g. the metallicity) becomes challenging and thus extracting stellar parameters such as the stellar mass is more difficult. 
%less stellar spectral lines
%hard to get chemical composition because fast rotation blend stellar spectral lines
%As a result, calculating stellar cross-correlation functions becomes challenging, which in turn reduces the precision of the derived stellar parameters.
Figure~\ref{fig:deltaKp} clearly illustrates the variation in the computed $\Delta\mathrm{K_p}$ as function of the stellar mass, but using the stellar reflex motion obtained from the radial velocity fitting in all four scenarios presented. Vertical lines mark the $\Delta\mathrm{K_p}$ values predicted by GCMs. We can surely claim that all computed planet's orbital velocity are consistent with either weak-drag or drag-free models, while the strong-drag configuration is ruled out, regardless of the model. These results are fully consistent with~\cite{wardenier_phase-resolving_2024}.

%Still, within the range of reported stellar masses, our $\Delta\mathrm{K_p}$ values remain in good agreement with the drag-free models discussed by~\cite{wardenier_phase-resolving_2024}, supporting the scenario of fast atmospheric dynamics and limited radiative damping in \hbox{WASP-121b}'s upper atmosphere.

%All detection signatures are consistent with a drag-free atmosphere at the altitude accessible with our observations, as predicted by global circulation models in~\cite{wardenier_phase-resolving_2024}. To further strengthen this interpretation, we calculate the $\mathrm{K_p}$ variation as $\Delta\mathrm{K_p=K_{p}^{retrieved}-K_{p}^{orbital}\simeq-17~km~s^{-1}}$, which closely matches predictions from circulation models incorporating no-drag conditions. %(with drag timescales $\tau \sim 10^4$s). 
%Enhanced drag slows atmospheric winds, leading to more vertically uniform temperature structures due to less efficient heat redistribution. Strong-drag models suppress equatorial jets and instead support dominant day-to-night winds. This circulation pattern is consistent with the Fe-traced dynamics reported by~\cite{seidel+2025_w121b_dynamics_nature} at millibar pressures, which aligns with the pressure levels probed by our observations.
%However, we must caution that the computation of the $\Delta\mathrm{K_p}$ strictly depends on the assumed stellar mass. We may thus compute different values which lead to different interpretations in the context of global circulation models.

\begin{table}[!h]
\caption{Mass reference. $\Delta\mathrm{K_p}$ computed assuming different stellar masses by keeping fixed the stellar reflex motion.}
\centering
\begin{tabular}{l c c}
\hline
\hline
Reference & $\mathrm{M_\star [M_\odot]}$ & $\Delta\mathrm{K_p~[km~s^{-1}]}$ \\
\hline
\cite{delrez_wasp-121_2016} & 1.35$\pm$0.08 & -14$\pm$5 \\
\cite{borsa_atmospheric_2021} & 1.38$\pm$0.02 & -15$\pm$3 \\
\cite{sing_absolute_2024}   & 1.33$\pm$0.02 & -13$\pm$3 \\
\cite{prinoth+2025_W121b_4UT} & 1.42$\pm$0.03 & -18$\pm$3\\
\hline
\hline
\end{tabular}
\label{tab:delta_kp_values}
\end{table}

%\subsection{Global Circulation Modelling??}
%Joost Wardenier. Provide the deltaKp that is predicted by different global circulation models at the spectral resolution of the bandpasses used in this work.

%--------------------------------------------------------------------
\section{Conclusions}
%Future perspectives: \hbox{WASP-121b} will be observed by PLATO in LOPS2 (LOng Pointing South field of view, see/cite Nascimbeni et al. January 2025).
In this work, we present an atmospheric retrieval analysis of the ultra-hot Jupiter \hbox{WASP-121b} by combining high-resolution transmission spectra from HARPS, NIRPS, and CRIRES+ (both K and H photometric bands). Leveraging the large high-resolution dataset, we constrain the planet's atmospheric composition and orbital dynamics through a free retrieval approach using the \texttt{SCARLET} code.

We report the retrieved relative chemical abundances of key species, including Fe, V, CO, H$_2$O, OH, TiO, H$^-$, and e$^-$. While absolute abundances remain highly model-dependent and less informative, abundance ratios offer a more robust window into the atmospheric processes and underlying chemistry. Our results indicate that the atmosphere of \hbox{WASP-121b} is broadly consistent with a solar metallicity and a solar C/O ratio. This suggests that chemical equilibrium remains largely preserved at the pressure levels probed ($\sim$10$^{-4}$–10$^{-3}$ bar), with no strong evidence for substantial alteration from disequilibrium processes such as rainout, photochemistry, or deep vertical mixing.

Notably, we constrain the $\log(\mathrm{H_2O/CO})$ ratio to be 1 dex sub-solar compared to chemical equilibrium predictions from \texttt{FastChem}, a result consistent with partial water dissociation at high temperatures. Despite the lack of a direct cross-correlation detection of H$_2$O, the retrieval favours the inclusion of water vapour to reproduce the observed spectra. The retrieved temperature of $2\,861^{+396}_{-418}$~K %confirms \hbox{WASP-121b}’s as an ultra-hot Jupiter and
reinforces the expectation of thermal dissociation affecting volatile species, such as water.

The $\log(\mathrm{CO/Fe})$ and $\log(\mathrm{V/Fe})$ ratios are in agreement with solar values, indicating that refractory-to-volatile balances are maintained and that CO remains stable against dissociation, consistent with its high bond energy. These findings reinforce the utility of CO as a chemically robust reference molecule in high-temperature atmospheres. In contrast, $\log(\mathrm{H^-/Fe})$ emerges as a potentially sensitive tracer of thermal ionisation. The retrieval returns a value 1 dex elevated (but still within 1$\sigma$) compared to solar predictions, hinting at either an elevated abundance of hydride continuum opacity but likely to compensate for missing opacity contributions in the model assumption, or a depleted Fe content due to condensation. However, the retrieval fails to constrain the electron density, limiting our ability to distinguish between enhanced H$^-$ opacity and Fe depletion through condensation.

Intriguingly, the cross-correlation signals for Fe, V, and CO appear blueshifted along the radial velocity axis and peak at a lower orbital velocity than the reference one in $\mathrm{K_p}$-$\mathrm{V_{sys}}$ space, suggesting the presence of equatorial winds and rotational broadening. These results are potentially indicative of drag-free or weak-drag atmospheric circulation as predicted by global circulation models. These models support a scenario in which CO probes higher-altitude layers subject to faster wind speeds, while Fe and V trace deeper regions where rotation dominates.

We also re-examine the planet’s orbital parameters using the largest radial velocity dataset to date. We report on a global fit to radial velocity measurements (from CORALIE, HARPS, NIRPS, and ESPRESSO observations) and photometric light curves from TESS in order to refine the planetary orbital parameters. Indeed, thanks to the broad radial velocity dataset, the orbital parameters of the system are improved. 
By comparing the derived orbital velocity  from
both the radial velocity analysis and the atmospheric retrieval, we estimate the velocity offset, $\Delta \mathrm{K_p}$, and find it
is non-zero, but consistent with predictions from either drag-free or weak-drag 3D global circulation models, while cautioning the non-negligible dependence on the stellar mass assumed.

Overall, our results provide a chemically and dynamically coherent picture of \hbox{WASP-121b}’s atmosphere. The planet's bulk composition appears largely unaltered by disequilibrium processes, while local deviations—such as water dissociation and thermal ionisation—highlight the complexity of atmospheric processes in ultra-hot environments. This study demonstrates the power of combining multi-instrument high-resolution datasets with retrieval techniques to simultaneously probe chemical composition and atmospheric dynamics. 

Future work will benefit from improved opacity databases—especially for TiO—and more physically complete models incorporating vertical mixing, photochemistry, and cloud condensation. The integration of low- and high-resolution datasets within a joint retrieval framework will be essential to further refine molecular abundance constraints. %The upcoming ESA mission PLATO, as discussed in~\cite{Nascimbeni+2025_PLATO_field}, will play a crucial role by providing ultra-precise stellar parameters and improved transit timing measurements. This will significantly reduce uncertainties on host star properties and planetary radii, which currently limit the precision of absolute abundance determinations in retrieval studies.

% --------------------------------------

\begin{acknowledgements}
%DE and MS  acknowledge support from the Swiss National Science Foundation for project 200021\_200726. The authors acknowledge the financial support of the SNSF.\\
This work has been carried out within the framework of the NCCR PlanetS supported by the Swiss National Science Foundation under grants 51NF40\_182901 and 51NF40\_205606.\\
RA  acknowledges the Swiss National Science Foundation (SNSF) support under the Post-Doc Mobility grant P500PT\_222212 and the support of the Institut Trottier de Recherche sur les Exoplan\`etes (IREx).\\
RA, JPW, \'EA, FBa, BB, NJC, RD, LMa \& CC  acknowledge the financial support of the FRQ-NT through the Centre de recherche en astrophysique du Qu\'ebec as well as the support from the Trottier Family Foundation and the Trottier Institute for Research on Exoplanets.\\
ML, HC \& BA  acknowledge support of the Swiss National Science Foundation under grant number PCEFP2\_194576.\\
DE and MS  acknowledge support from the Swiss National Science Foundation for project 200021\_200726. \\
VV, SP, DE, \& MS acknowledge the financial support of the SNSF.\\
NN, JIGH, RR, ASM \& AKS  acknowledge financial support from the Spanish Ministry of Science, Innovation and Universities (MICIU) projects PID2020-117493GB-I00 and PID2023-149982NB-I00.\\
NN  acknowledges financial support by Light Bridges S.L, Las Palmas de Gran Canaria.\\
NN acknowledges funding from Light Bridges for the Doctoral Thesis "Habitable Earth-like planets with ESPRESSO and NIRPS", in cooperation with the Instituto de Astrof\'isica de Canarias, and the use of Indefeasible Computer Rights (ICR) being commissioned at the ASTRO POC project in the Island of Tenerife, Canary Islands (Spain). The ICR-ASTRONOMY used for his research was provided by Light Bridges in cooperation with Hewlett Packard Enterprise (HPE).\\
XDu  acknowledges the support from the European Research Council (ERC) under the European Union’s Horizon 2020 research and innovation programme (grant agreement SCORE No 851555) and from the Swiss National Science Foundation under the grant SPECTRE (No 200021\_215200).\\
This work was supported by grants from eSSENCE (grant number eSSENCE@LU 9:3), the Swedish National Research Council (project number 2023 05307), The Crafoord foundation and the Royal Physiographic Society of Lund, through The Fund of the Walter Gyllenberg Foundation."\\
\'EA, FBa, RD \& LMa  acknowledges support from Canada Foundation for Innovation (CFI) program, the Universit\'e de Montr\'eal and Universit\'e Laval, the Canada Economic Development (CED) program and the Ministere of Economy, Innovation and Energy (MEIE).\\
SCB, ED-M, NCS \& ARCS  acknowledge the support from FCT - Funda\c{c}\~ao para a Ci\^encia e a Tecnologia through national funds by these grants: UIDB/04434/2020, UIDP/04434/2020.\\
SCB   acknowledges the support from Funda\c{c}\~ao para a Ci\^encia e Tecnologia (FCT) in the form of a work contract through the Scientific Employment Incentive program with reference 2023.06687.CEECIND and DOI \href{https://doi.org/10.54499/2023.06687.CEECIND/CP2839/CT0002}{10.54499/2023.06687.CEECIND/CP2839/CT0002.}\\
XB, XDe \& TF  acknowledge funding from the French ANR under contract number ANR\-24\-CE49\-3397 (ORVET), and the French National Research Agency in the framework of the Investissements d'Avenir program (ANR-15-IDEX-02), through the funding of the ``Origin of Life" project of the Grenoble-Alpes University.\\
The Board of Observational and Instrumental Astronomy (NAOS) at the Federal University of Rio Grande do Norte's research activities are supported by continuous grants from the Brazilian funding agency CNPq. This study was partially funded by the Coordena\c{c}\~ao de Aperfei\c{c}oamento de Pessoal de N\'ivel Superior—Brasil (CAPES) — Finance Code 001 and the CAPES-Print program.\\
BLCM  acknowledge CAPES postdoctoral fellowships.\\
BLCM  acknowledges CNPq research fellowships (Grant No. 305804/2022-7).\\
NBC  acknowledges support from an NSERC Discovery Grant, a Canada Research Chair, and an Arthur B. McDonald Fellowship, and thanks the Trottier Space Institute for its financial support and dynamic intellectual environment.\\
JRM  acknowledges CNPq research fellowships (Grant No. 308928/2019-9).\\
ED-M  further acknowledges the support from FCT through Stimulus FCT contract 2021.01294.CEECIND. ED-M  acknowledges the support by the Ram\'on y Cajal contract RyC2022-035854-I funded by MICIU/AEI/10.13039/501100011033 and by ESF+.\\
ICL  acknowledges CNPq research fellowships (Grant No. 313103/2022-4).\\
CMo  acknowledges the funding from the Swiss National Science Foundation under grant 200021\_204847 “PlanetsInTime”.\\
Co-funded by the European Union (ERC, FIERCE, 101052347). Views and opinions expressed are however those of the author(s) only and do not necessarily reflect those of the European Union or the European Research Council. Neither the European Union nor the granting authority can be held responsible for them.\\
GAW is supported by a Discovery Grant from the Natural Sciences and Engineering Research Council (NSERC) of Canada.\\
0\\
KAM  acknowledges support from the Swiss National Science Foundation (SNSF) under the Postdoc Mobility grant P500PT\_230225.\\
ARCS  acknowledges the support from Funda\c{c}ao para a Ci\^encia e a Tecnologia (FCT) through the fellowship 2021.07856.BD.\\
BP  acknowledges financial support from the Walter Gyllenberg Foundation.\\
AKS  acknowledges financial support from La Caixa Foundation (ID 100010434) under the grant LCF/BQ/DI23/11990071.\\
BT  acknowledges the financial support from the Wenner-Gren Foundation (WGF2022-0041).\\
AP acknowledges support from the Unidad de Excelencia María de Maeztu CEX2020-001058-M programme and from the Generalitat de Catalunya/CERCA.
\end{acknowledgements}

% --------------------------------------

\bibliographystyle{aa} % style aa.bst
\bibliography{references}

% --------------------------------------

\onecolumn
\begin{appendix}
\section{Detrending steps}
\label{section:appendix_deterending_steps}
\begin{figure*}[ht!]
    \centering
        \begin{minipage}[b]{0.49\textwidth}  \centering\includegraphics[width=1.2\linewidth]{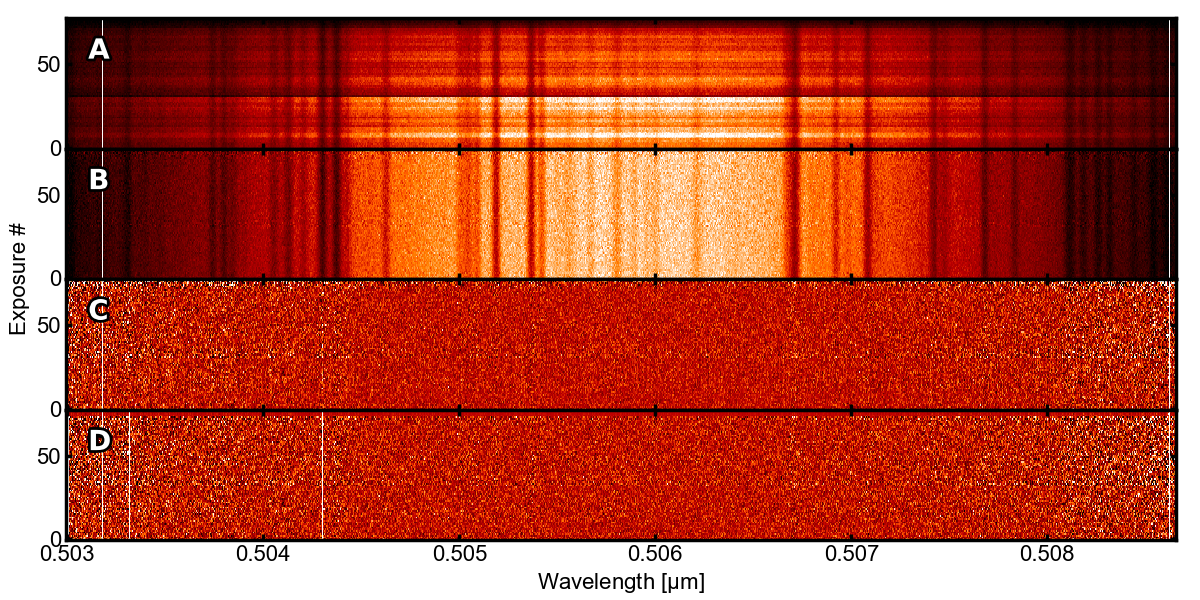}
    \caption*{(a) Example of a  reduced HARPS transit time-series gathered on January 5$^\text{th}$, 2025.}
    \end{minipage}
    %\hfill

    \begin{minipage}[b]{0.49\textwidth}
        \centering
        \includegraphics[width=1.2\linewidth]{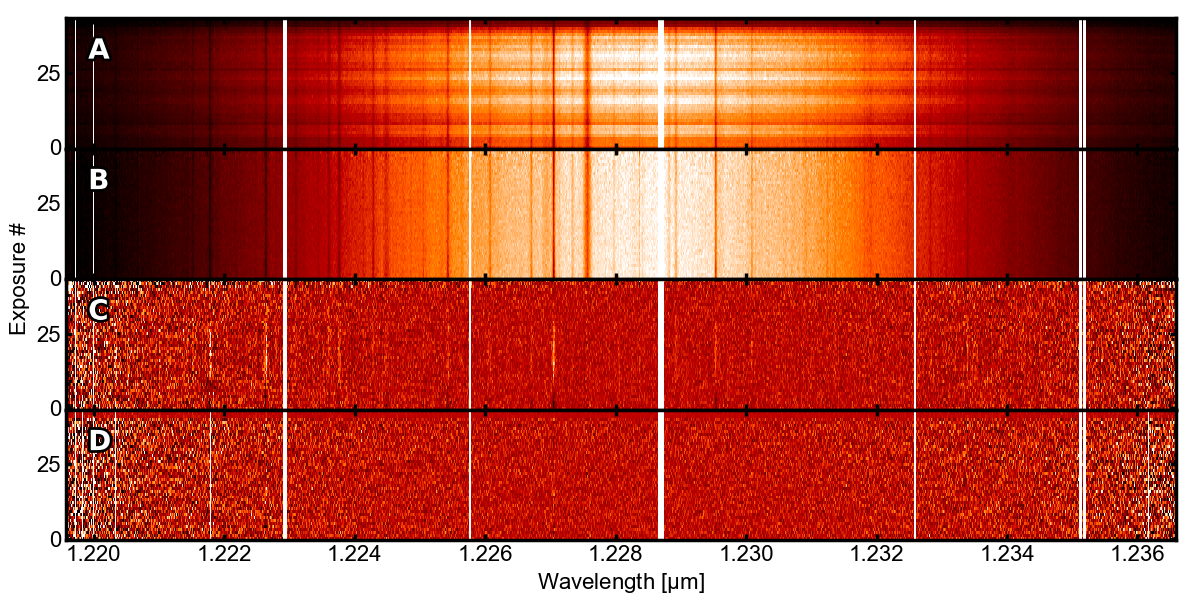}
        \caption*{(b)Example of a  reduced NIRPS transit time-series gathered on January 5$^\text{th}$, 2025.}
    \end{minipage}
    %hfill
    
    \begin{minipage}[b]{0.49\textwidth}
        \centering
        \includegraphics[width=1.2\linewidth]{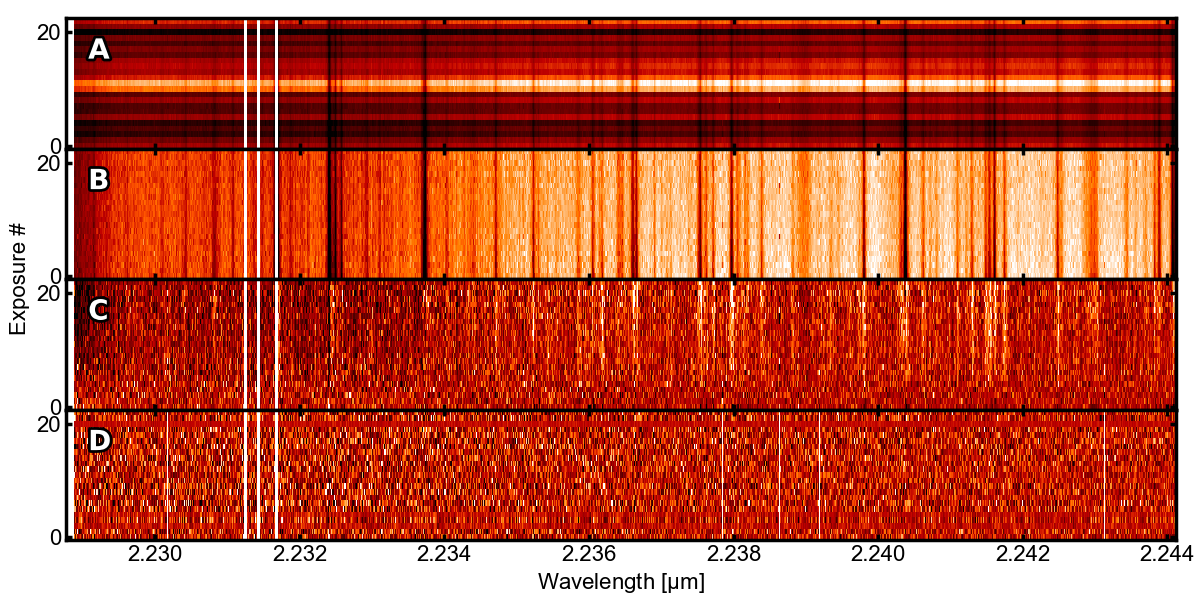}
        \caption*{(c) Example of a  reduced CRIRES+ transit time-series gathered on December 21$^\text{st}$, 2021.}
    \end{minipage}

    \caption{Data detrending procedure for an example spectral order (exposures versus wavelengths) from HARPS (panel a), NIRPS (panel b), and CRIRES+ (panel c).
Each panel shows the following steps: A — the raw time-series of a single spectral order during the transit; B — after continuum normalization; C — after dividing out the median out-of-transit spectrum; D — final residuals after removing stellar and telluric spectral lines using PCA.}
    \label{fig:appendix_detrending_procedure}
\end{figure*}

\newpage
\section{Correction for stellar effects}
%\afterpage{%
\begin{figure*}[h!]
    \centering
    \begin{minipage}{0.49\textwidth}
        \centering
        \includegraphics[width=\linewidth]{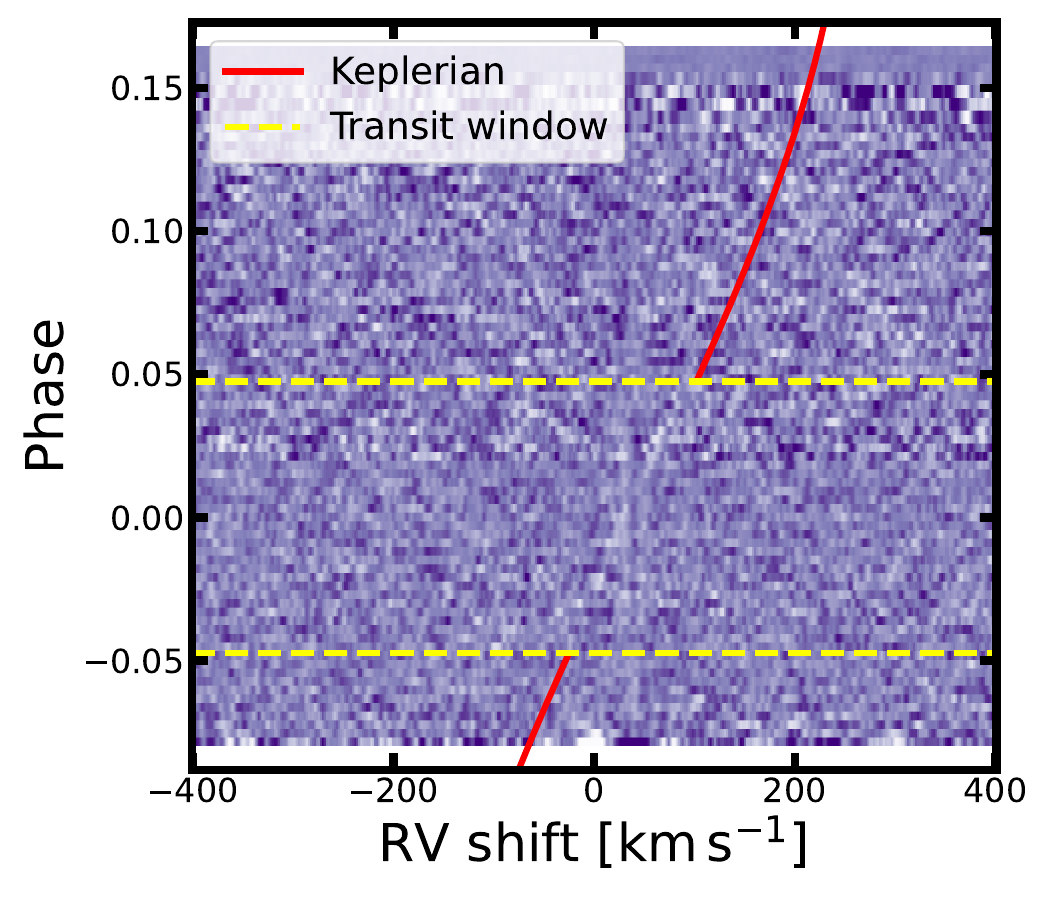}
        \caption*{(a) HARPS data not masked for the Doppler shadow.}
    \end{minipage}
    \begin{minipage}{0.49\textwidth}
        \centering
        \includegraphics[width=\linewidth]{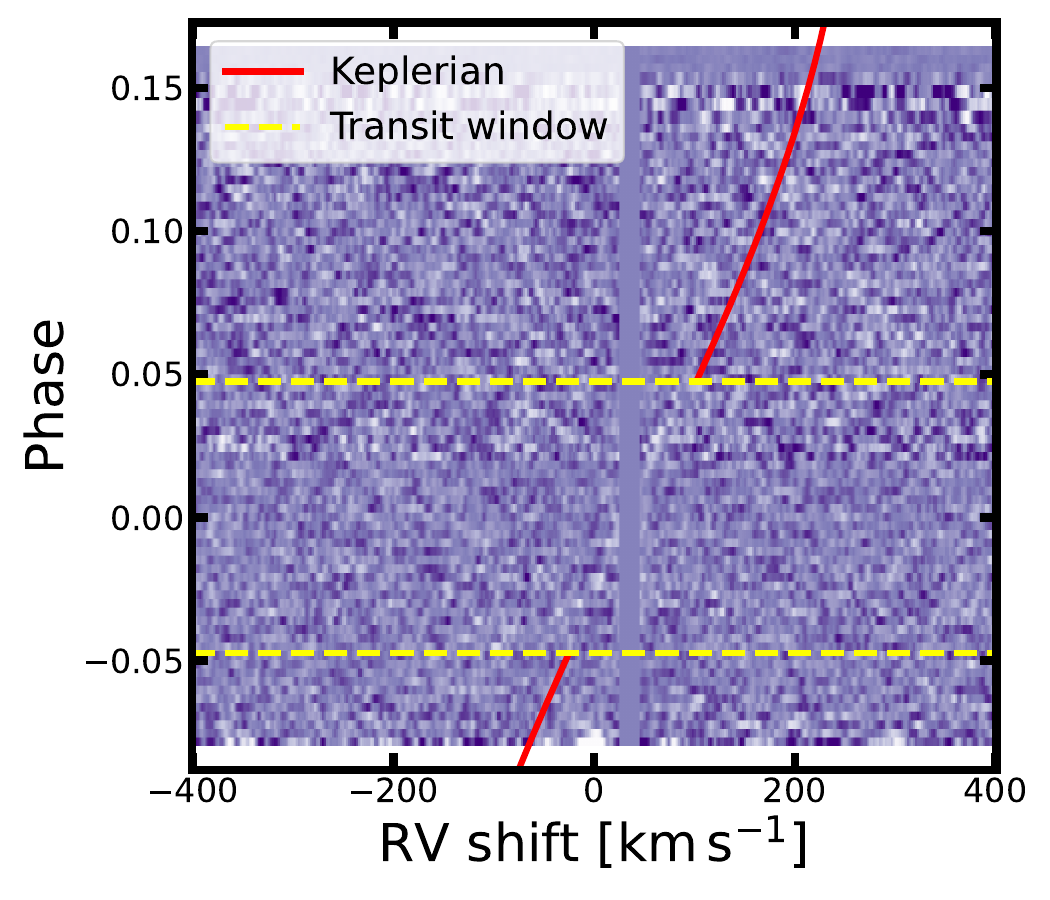}
        \caption*{(b) HARPS data masked for the Doppler shadow around the reference systemic velocity ($\simeq38~\mathrm{km~s^{-1}}$).}
    \end{minipage}
    \caption{Iron cross-correlation trail maps for HARPS optical data exhibiting non-negligible stellar contamination (i.e. Rossiter–McLaughlin effect) at the systemic velocity expected from~\cite{delrez_wasp-121_2016}. Dashed yellow lines define the transit window. The solid red line illustrates the Keplerian curve (i.e. the planet's orbital motion). Panel (a) shows the HARPS optical data not corrected for the stellar contamination as Doppler shadow, while panel (b) shows the Doppler shadow masked (corresponding flux values are set to NaN and discarded).}
    \label{fig:appendix-DopplerShadow}
\end{figure*}
%}

\end{appendix}

% --------------------------------------

\end{document}